 \documentclass[final,5p,times,twocolumn]{elsarticle}

\usepackage{lineno}

\usepackage{booktabs}

\usepackage{tabularx}

\usepackage{multirow}
\usepackage{verbatim}
\usepackage{cleveref}

\usepackage{soul}
\usepackage{amsfonts}
\usepackage{csquotes}
\usepackage{amssymb}
\usepackage{amsmath}
\usepackage[export]{adjustbox}

\usepackage{makecell}
\usepackage{xcolor}
\usepackage[normalem]{ulem}

\usepackage{subcaption}

\newcommand\ore[1]{\noindent{\color{black} {#1}}}
\newcommand\reduline{\bgroup\markoverwith
	{\textcolor{magenta}{\rule[0.3ex]{2pt}{2.4pt}}}\ULon}

\graphicspath{{../pdf/}{../jpeg/}{figures/}}
\usepackage{array}
\usepackage{url}
\journal{Computer Methods and Programs in Biomedicine}
\makeatletter
\def\ps@pprintTitle{%
	\let\@oddhead\@empty
	\let\@evenhead\@empty
	\def\@oddfoot{\textcopyright~2020. This work is licensed under a CC BY-NC-ND 4.0 International License:\\ http://creativecommons.org/licenses/by-nc-nd/4.0/}%
	\let\@evenfoot\@oddfoot}
\makeatother

\begin{document}
	
	\begin{frontmatter}
		
		
		\title{Accelerating B-spline {Interpolation on GPUs}: Application to Medical Image Registration}
		


		% andrea orcid \textsuperscript{0000-0003-4764-5685}
		% Orestis~Zachariadis\textsuperscript{0000-0002-7433-8448}
		% Joaqu\'in~Olivares\textsuperscript{0000-0003-2584-5491}
		
		\author[1]{Orestis~Zachariadis\corref{*}}
		\cortext[*]{Corresponding author}
		\ead{orestis.zachariadis@uco.es}
		
		\author[2,3]{Andrea~Teatini}
		\ead{andre_tea@outlook.com}
		\author[1]{Nitin Satpute}
		\author[4]{Juan~G\'omez-Luna}
		\author[4]{Onur Mutlu}
		\author[2,3]{Ole~Jakob~Elle}
		\author[1]{Joaqu\'in~Olivares}
		
		\address[1]{Department of Electronics and Computer Engineering, Universidad de Cordoba, C\'ordoba, Spain}
		\address[2]{The Intervention Centre, Oslo University Hospital - Rikshospitalet, Oslo, Norway}
		\address[3]{Department of Informatics, University of Oslo, Oslo, Norway}
		\address[4]{Department of Computer Science, ETH Zurich, Zurich, Switzerland}

		%\author{Orestis~Zachariadis, Andrea~Teatini, , , , Ole Jakob Elle, % <-this % stops a space
		% 	\thanks{O. Zachariadis,  N. Satpute, J. Olivares are with the Department of Electronics and Computer Engineering, Universidad de Cordoba, C\'ordoba, Spain. e-mail: \{orestis.zachariadis, el2sasan, olivares\}@uco.es.}% <-this % stops a space
		% 	\thanks{J. G\'omez-Luna is with the Department of Computer Science, ETH Zurich, Zurich, Switzerland.}% <-this % stops a space
		% 	\thanks{Andrea Teatini and Ole Jakob Elle are affiliated with The Intervention Center, Oslo University Hospital, and with the Department of Informatics, University of Oslo, Oslo, Norway. e-mail: andrea.teatini@rr-research.no, oelle@ous-hf.no}% <-this % stops a space
		% 	\thanks{Manuscript received 31 of July, 2017.}
		%}

		\begin{abstract}
			%\boldmath
			%Background and Objective:
			\paragraph{Background and Objective}
			B-spline interpolation (BSI) is a popular technique in the context of medical imaging due to its adaptability and robustness in 3D object modeling. A field that utilizes BSI is Image Guided Surgery (IGS). IGS provides navigation using medical images, which can be segmented and reconstructed into 3D models, often through BSI. {Image} registration tasks also use BSI %, for example, 
			to transform medical imaging data collected before the surgery and intra-operative data collected during the surgery into a common coordinate space.
			However, {such} IGS tasks are computationally demanding, especially when applied to 3D medical images, due to the complexity and amount of data involved. Therefore, optimization of {IGS} algorithms is greatly desirable, for example, to perform {image} registration tasks intra-operatively and to enable real-time applications. A traditional CPU does not have sufficient computing power to achieve these goals and, thus, it is preferable to rely on GPUs. 
			In this paper, we introduce a novel GPU implementation of BSI to accelerate the calculation of the deformation field in non-rigid {image} registration algorithms.	
			
			%Methods
			\paragraph{Methods}
			Our BSI implementation on GPUs minimizes the data that needs to be moved between memory and processing cores during loading of the input grid, and leverages the large on-chip GPU register file for reuse of input values.
			Moreover, we re-formulate our method as trilinear interpolations to reduce computational complexity and increase accuracy.
			To provide pre-clinical validation of our method and {demonstrate} its {benefits in} medical applications, we integrate our improved BSI into a registration workflow for compensation of liver deformation (caused by pneumoperitoneum, i.e., inflation of the abdomen) and evaluate its performance.
			% 	\reduline{	We test registration in terms of accuracy and speed on pre-operative and intra-operative MRI scans of the liver of a porcine model and CT scans of a human liver \ore{model (phantom)}.}
			
			%Results
			\paragraph{Results}
			%\reduline{Our improved BSI approach improves accuracy by up to 3300$\times$ and performance by up to 7$\times$ compared to three state-of-the-art GPU implementations.} 
			{Our approach improves the performance of BSI by an average of 6.5$\times$ and interpolation accuracy by 2$\times$ compared to three state-of-the-art GPU implementations.} Through pre-clinical validation, we {demonstrate} that our optimized interpolation accelerates a non-rigid {image} registration algorithm, which is based on the Free Form Deformation (FFD) method, by up to 34\%.
			
			%Conclusion
			\paragraph{Conclusion}
			Our study shows that we can achieve significant performance and accuracy gains with	our novel parallelization scheme that makes {effective} use of the GPU resources. We show that our method improves the performance of {real} medical imaging registration applications {used in practice today}.
		\end{abstract}
		
		
	\end{frontmatter}
	
%	\linenumbers

	\section{Introduction}
	
	Image Guided Surgery (IGS) aims to provide surgeons with navigation capabilities to perform safer surgeries through better visualization~\cite{Bartoli2012}. IGS is created by combining medical images, such as Computed Tomography (CT) {or} Magnetic Resonance Imaging (MRI) \cite{Bernhardt2017}, with surgical instrument tracking technologies~\cite{teatini2018assessment}. However, the accuracy of image guided surgery is often undermined by organ deformations, especially in soft tissue surgeries. 
	These %movements 
	{deformations} are difficult to account for due to their non-linear behaviour. Non-rigid registration is a technique that has been developed to reproduce and model {such} non-linear deformations \cite{sotiras2013deformable}. 
	
	Non-rigid registration through Free Form Deformation (FFD) \cite{Rueckertetal.1999}, based on cubic B-spline interpolation (BSI)~\cite{Rueckertetal.1999,Ellingwood2016}, is a %suitable solution 
	{state-of-the-art technique} for non-rigid registration.
	FFD works by manipulating a grid of control points. The shape of a 3D object (e.g., an organ) underlying the control points can be changed by using a smooth and $C^2$ continuous transform (i.e., continuous up to second order derivatives).
	%\jgl{What is $C^2$? Clarify}
	FFD uses BSI in the calculation of the deformation field.
	
	BSI is one of the most computationally demanding parts of FFD \cite{Modat2010}. %\reduline{Registration is a computationally demanding task, where BSI takes approximately 30\% of the total execution time of FFD, according to our experiments}. 
	Graphics Processing Units (GPUs) can help achieve the real-time requirements of IGS, namely FFD, as they offer massive {computational} performance in comparison to Central Processing Units (CPUs). GPUs deploy thousands of execution threads, which operate on large batches of data. GPUs provide higher throughput and power-efficiency than CPUs on multithreaded workloads~\cite{ProgrGuide}. The performance of medical imaging applications benefits significantly from GPUs \cite{smistad2015medical,gai2013more,stone_accelerating_2008, wang2018survey, kalaiselvi2017survey,palomar2018high,nitin2020}.
	% 	\jgl{Include a sentence saying that GPUs have shown impressive performance improvements in medical image applications. Cite Smistad's survey and more papers (include citations to Wen-mei's papers on MRI on GPU -- find them here: http://impact.crhc.illinois.edu/publications.aspx}
	%Therefore, GPUs can improve the performance of BSI significantly \cite{Ruijters2008,sigg2005fast}.
	
	For these reasons, several authors have used GPUs for BSI~\cite{Ellingwood2016, sigg2005fast, Ruijters2008, Du2016, Shackleford2010}.
	Sigg et al. \cite{sigg2005fast} and Ruijters et al. \cite{Ruijters2008} achieve a substantial reduction in the number of input samples by representing the weighted sums as trilinear interpolations. More recently, Ellingwood et al. \cite{Ellingwood2016} and Du et al. \cite{Du2016} use GPU implementations of BSI to improve the performance of {image} registration. 
	They improve input sample loading by aligning the control grid with the voxel grid of the volume~\cite{Ellingwood2016, Du2016, Shackleford2010}.
	However, all these works suffer from the intensive \emph{data movement} of a large number of input samples between the memory and the GPU, which is the main {performance} bottleneck of BSI implementations on a GPU \cite{sigg2005fast}.
	
	%\jgl{State clearly what is \textbf{the problem} of previous works. State what is \textbf{our goal} in this work.}
	
	Our goal in this work is to accelerate BSI on {GPUs} by alleviating the data movement bottleneck with optimization techniques that enable a more efficient use of the on-chip memory resources. 
	To this end, we propose a GPU implementation of BSI with three key optimizations: 
	a) a new workload partitioning scheme for GPU execution threads that reduces the number of memory accesses, 
	b) a %register-only 
	{register-tiling} approach that keeps input data close to the execution units, and 
	c) the replacement of weighted summation with linear interpolations, which reduces the computational load and increases the accuracy. 
	
	In order to show how our approach %impacts 
	affects the performance and accuracy of {image} registration in a realistic scenario, we integrate %it 
	our technique (\ore{publicly available}\footnote{\url{https://github.com/oresths/niftyreg_bsi}}) to the FFD registration of NiftyReg~\cite{Modat2010}. NiftyReg is a lightweight medical image registration library. Recent works \cite{peterlik2018fast, lee2015evaluation}
	%\jgl{Recent works is 1 work? Add more references.}
	use NiftyReg as a reference for registration.

	We complete our study with a pre-clinical evaluation of our method.
	We use FFD with our GPU-accelerated BSI on 1) CT scans of patient-specific liver phantom, and 2) MRI scans of a porcine liver model to compensate for a non-rigid soft tissue deformation caused by pneumoperitoneum. Pneumoperitoneum is a surgical procedure to inflate the patient's abdomen, which is necessary for any abdominal laparoscopic surgery. Pneumoperitoneum, however, deforms the shape of the organs~\cite{Heiselman2018,Johnsen2015}. To account for this deformation, we capture new images during the surgery (intra-operative) and use non-rigid image registration to match them with images before pneumoperitoneum (pre-operative images). %Non-rigid registration for pneumoperitoneum was computed through our BSI implementation and state-of-art algorithms, deeming an increase of performance using our method with the same accuracy as the other methods.
	{We compute non-rigid image registration for pneumoperitoneum with state-of-art implementations and with our BSI implementation. Using our implementation results in a performance increase with the same accuracy as using the state-of-the-art implementations}.
	%\jgl{Please include here the main conclusions/summary of the pre-clinical evaluation.}

	\section{Background}
	In this section, we first introduce the foundations of B-spline interpolation. 
	Since our GPU implementation of BSI is specific to 3D medical images (CT, MRI, or US volumes), formulations and analysis focus on the 3D case.
	Second, we review two state-of-the-art implementations of BSI on GPUs.
	
	\subsection{B-spline interpolation theory} \label{sec:bspline_theory}
	
\begin{figure}[h]
		\centering
		\includegraphics[width=0.8\linewidth]{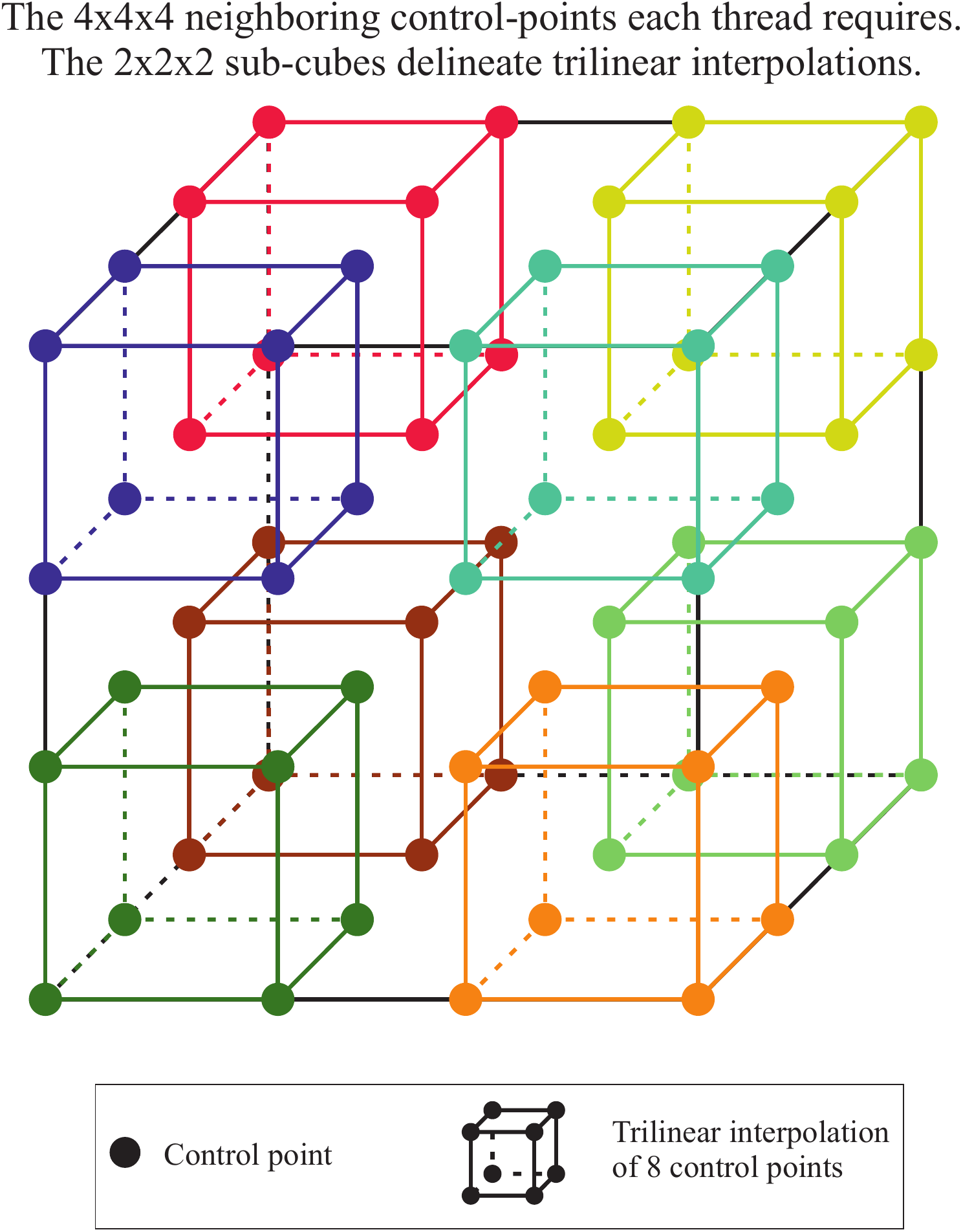}
		\caption{The cube of $4 \times 4 \times 4$ control points that affect a voxel/tile in a 3D control point grid. Smaller cubes depict the grouping in trilinear interpolations.}
		\label{fig:lerp}
	\end{figure}

	We introduce B-spline interpolation for 3D images, i.e., the domain of the image volume is in the $ x, y, z $ coordinate space. 
	As Equation~\ref{eq:sum} shows~\cite{Rueckertetal.1999,Ellingwood2016}, the BSI transformation of FFD for each voxel (i.e., each interpolated point of FFD) with coordinates $ x, y, z $ is $T(x, y, z)$. The BSI transformation is a function of control points $ \phi_{i,j,k} $, which are {arranged into a grid of dimensions $ n_{x} \times n_{y} \times n_{z} $.
		%We denote the elements of a $ n_{x} \times n_{y} \times n_{z} $ control point grid as $ \phi_{i,j,k} $. 
		The control point grid is} uniformly spaced, with $\delta_{x}$, $\delta_{y}$, and $\delta_{z}$ being the spacing {(in voxels)} in the three dimensions.
	%\ore{We denote the uniform spacing in voxels (where voxels refer to the interpolated points of FFD) of neighboring control points as $ \delta_{x} \times \delta_{y} \times \delta_{z} $.} \ore{Four control points in each of the $ x, y, z $ directions affect each voxel. $ T(x, y, z) $ calculates the BSI transformation of FFD} as \cite{Rueckertetal.1999,Ellingwood2016}:
	
	\begin{equation} \label{eq:sum}
	T(x, y, z)= \sum_{l=0}^{3}\sum_{m=0}^{3}\sum_{n=0}^{3}B_l(u)B_m(v)B_n(w)\phi_{i+l,j+m,k+n} 
	\end{equation}
	
	where $ i=\lfloor x/\delta_{x} \rfloor -1, j=\lfloor y/\delta_{y} \rfloor -1, k=\lfloor z/\delta_{z} \rfloor -1, u=x/\delta_{x} - \lfloor x/\delta_{x} \rfloor, v=y/\delta_{y} - \lfloor y/\delta_{y} \rfloor, w=z/\delta_{z} - \lfloor z/\delta_{z} \rfloor$, $B$ are the \ore{scalar} {B-spline coefficients \cite{Ruijters2008}} %\jgl{You have to explain how they are obtained or, at least, include a citation.} 
	and $ \phi $ are the control points. 
	Each voxel is affected by four control points in each dimension. Thus, in a 3D space, $4 \times 4 \times 4$ control points, forming a cube (see Figure \ref{fig:lerp}), affect the inner \emph{tile} of voxels.
	%\ore{We observe that $4 \times 4 \times 4$ control points, forming a cube (Figure \ref{fig:lerp}), affect each tile of voxels.} 
	In general, in N-dimensional images, $ 4^N $ control points affect each voxel.
	
	%\jgl{This expression is not clear yet. You have removed my comments without addressing them.}\ope{x,y,z are part of i,j,k. - I had missed j,k. l,m,n are the indices of the summation operator. I hope my new comment about the 4 control points addressed the index issue.}
	
	\subsubsection{Tiles} \label{ssec:tiles}
	
\begin{figure}[h]
		\centering
		\includegraphics[width=0.7\linewidth]{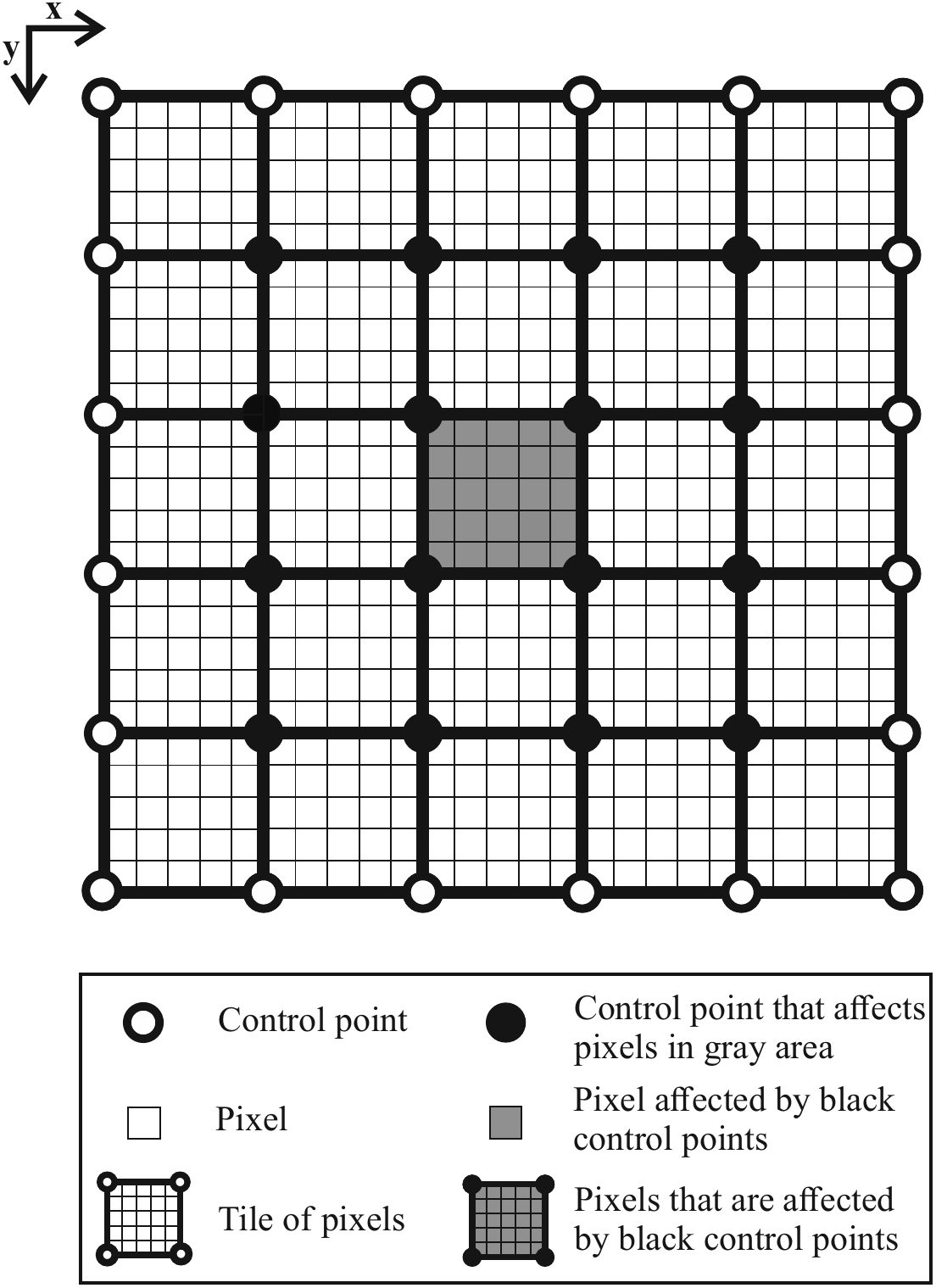}
		%		\caption{Tile properties of a 2D image \jgl{These are not "properties". These might be "sample tiles", "a 2D space divided into tiles", or something similar, not "tile properties".}}
		\caption{A 2D space divided into tiles.}
		\label{fig:tile}
	\end{figure}

	\emph{Tiles} are logical groups of voxels that share common properties. Based on Equation \ref{eq:sum}, we define tiles of $ \delta_{x} \times \delta_{y} \times \delta_{z} $ dimensions.
	%	 \jgl{This is confusing. You talk about tiles of $ \delta_{x} \times \delta_{y} \times \delta_{z} $ dimensions, but it is not clear how they relate to the $4 \times 4 \times 4$ control points. Btw, in Section 2.1 $ \delta_{x} \times \delta_{y} \times \delta_{z} $ are the uniform spacing in control grids? Not clear at all.}. 
	Figure \ref{fig:tile} illustrates a tile in a 2D example. We make two observations: 1) the \emph{same} control points, i.e., the ones surrounding the tile, affect all voxels inside the tile, and 2) control points of neighboring tiles overlap.
	
	%We utilize tiles in our proposed optimization scheme to significantly reduce memory traffic between off- and on-chip memory.
	From the implementation perspective, partitioning a volume into tiles is a way of exploiting data reuse (i.e., reuse of control points) in on-chip memories, when calculating the interpolated voxels. 
	Thus, \emph{tiling} saves memory traffic between off-chip and on-chip memories.

	\subsection{State-of-the-art GPU implementations of BSI} \label{sec:soa}
	This section introduces the two state-of-the-art BSI methods and their respective GPU implementations, which we use as comparison points for our work.
	
	\paragraph{Texture Hardware (TH)} \label{ssec:TH}
	Ruijters et al.~\cite{Ruijters2008,Ruijters2012} provide a texture hardware method for BSI. They base their method on the observation that the weighted additions of Equation \ref{eq:sum}
	%	\jgl{What weighted addition? Refer to Eq. 1}
	can be replaced by a linear interpolation \cite{sigg2005fast,Ruijters2008}.
	%	\jgl{Clarify why can be replaced.}
	Linear interpolations are well-suited for the GPU texture unit, that %encompasses 
	features a hardware interpolation unit. The hardware interpolation unit calculates the interpolation directly and it does not require separate accesses to off-chip \emph{global} memory of the GPU to load the input control points.
	Hardware interpolation is fast but it has two main drawbacks. 
	First, it has only 8 bits of accuracy \cite{ProgrGuide}, which limits the resolution of the interpolation. 
\ore{Second, the values that the hardware interpolation unit fetches from the off-chip memory are a function of the absolute position of each voxel. Therefore, TH cannot utilize custom caching schemes to aggregate data transfers for neighboring voxels (\ref{sec:mem_transf}).}
	%	 \jgl{Is this a drawback? Why caching something that does not have reuse? The results of interpolation are the output, no? Why caching them?}. 
	Texture Hardware BSI is included in an easy-to-use library by Ruijters et al.~\cite{Ruijters2012} and is used in recent works~\cite{andersson2016fast,carron2017maximum}.
	
	\paragraph{Thread per Voxel (TV)}
	%The basic mechanism of this approach is that each CUDA thread is assigned to a single element in a straightforward way, i.e., a thread for each voxel in the case of 3D images. 
	This method assigns one thread per image element, e.g., per voxel in the case of 3D images.
	%	\jgl{The method is assigning one "CUDA" thread? Can't you use TV method on a CPU and assign one CPU thread per voxel? You should first present the generic method (i.e., one thread per voxel) and then talk about the implementation for a particular hardware (i.e., one CUDA thread per voxel or one CPU thread per voxel)}.
	
	Ellingwood et al.~\cite{Ellingwood2016} present a GPU implementation of this method that applies tiling (Section~\ref{ssec:tiles}). They assign one or more \emph{thread blocks} to each tile, with one thread for each voxel of the tile. Tiling enables the reuse of control points, which are the same for the whole tile, by keeping them in the fast on-chip \emph{shared} memory.
	
	NiftyReg~\cite{Modat2010}, a lightweight open-source medical image registration library, also uses the thread per voxel method. 
	NiftyReg contains optimized implementations of BSI for both CPUs and GPUs. It is open-source and well-maintained, with competitive performance against other state-of-the-art implementations \cite{peterlik2018fast, lee2015evaluation}. The GPU implementation uses a simple, straightforward TV method, which does not take advantage of tiling. 
	The CPU implementation, however, exploits tiling by applying multi-core and vectorization optimizations. %minor utilization of tile properties.

	\section{Optimizing B-spline interpolation} \label{sec:impl}
	This section presents our GPU implementation of BSI, which follows a different approach to the state-of-the-art implementations (i.e., TH and TV).
	In our approach, we assign \emph{one thread per tile} of voxels, as we explain in Sections~\ref{ssec:overview} to~\ref{ssec:ttli}.
	In Section~\ref{ssec:cpu}, we introduce our implementations for CPU, which follow the GPU approach partially.
	
	%\subsection{Overview of our approach}
	\subsection{Overview of our GPU implementation of BSI} \label{ssec:overview}
	%\jgl{If the approach is applicable to both CPU and GPU, it must be presented in an architecture-agnostic way.} \ope{It works only partially on CPU, I make a table later.}
	
	Our GPU implementation of BSI is based on two key ideas.
	
	First, an entire tile of voxels is assigned to a single GPU thread (\emph{Thread per Tile}, TT), in contrast to the one-thread, one-voxel approach. 
	%\reduline{	Figure \ref{fig:TT} compares the thread assignment of TV and our approach. }
	%\jgl{You have to explain the figure. TV in Fig. 3 corresponds to Ellingwood, NiftyReg or both?}
	%\ope{I include the figure later, added more info in caption}ichi
	%With this \ore{thread} assignment, we minimize both the reads from off-chip memory by maximizing the overlap of the input control points, and the cache accesses by keeping the input points in registers and reusing them by many voxels. 
	This TT assignment takes advantage of tiling in both \ore{on-chip \emph{cache}} memory and registers: 1) tiling in \ore{cache} memory minimizes the reads from off-chip memory, by maximizing the overlap of input control points, and 2) tiling in registers minimizes the accesses to cache memory, by reusing the input control points for many voxels.
	
	Second, we replace the weighted sum of the basic formula of BSI with trilinear interpolations, %\ore{similar to TH.}
	in a similar way as TH does.
	%\jgl{How is this done? Why possible? Do you explain it later? Any reference? Clarify}.
	We calculate these trilinear interpolations using Fused Multiply-Add (FMA) instructions, which the GPU instruction set contains~\cite{ProgrGuide}. 
	FMA increases both accuracy and speed in regard to regular multiplication and addition instructions. 
	
	We give an in-depth description of our optimizations in the next sections.
	
	%	\jgl{Present what you are going to explain in the following subsections.}
	
	\subsection{Thread per Tile (TT)} \label{ssec:tt}
	
	%In the following paragraphs we describe the optimizations utilized in the proposed implementations. 
	In this section, we describe the optimization techniques that we deploy in our TT approach to BSI.
	We show how the input loading and register optimizations reduce memory accesses.
	
	\subsubsection{Input loading optimization} \label{sec:input_optimiz}
\begin{figure*}[h]
		\centering
		\captionsetup{justification=centering}
		\includegraphics[width=0.8\linewidth]{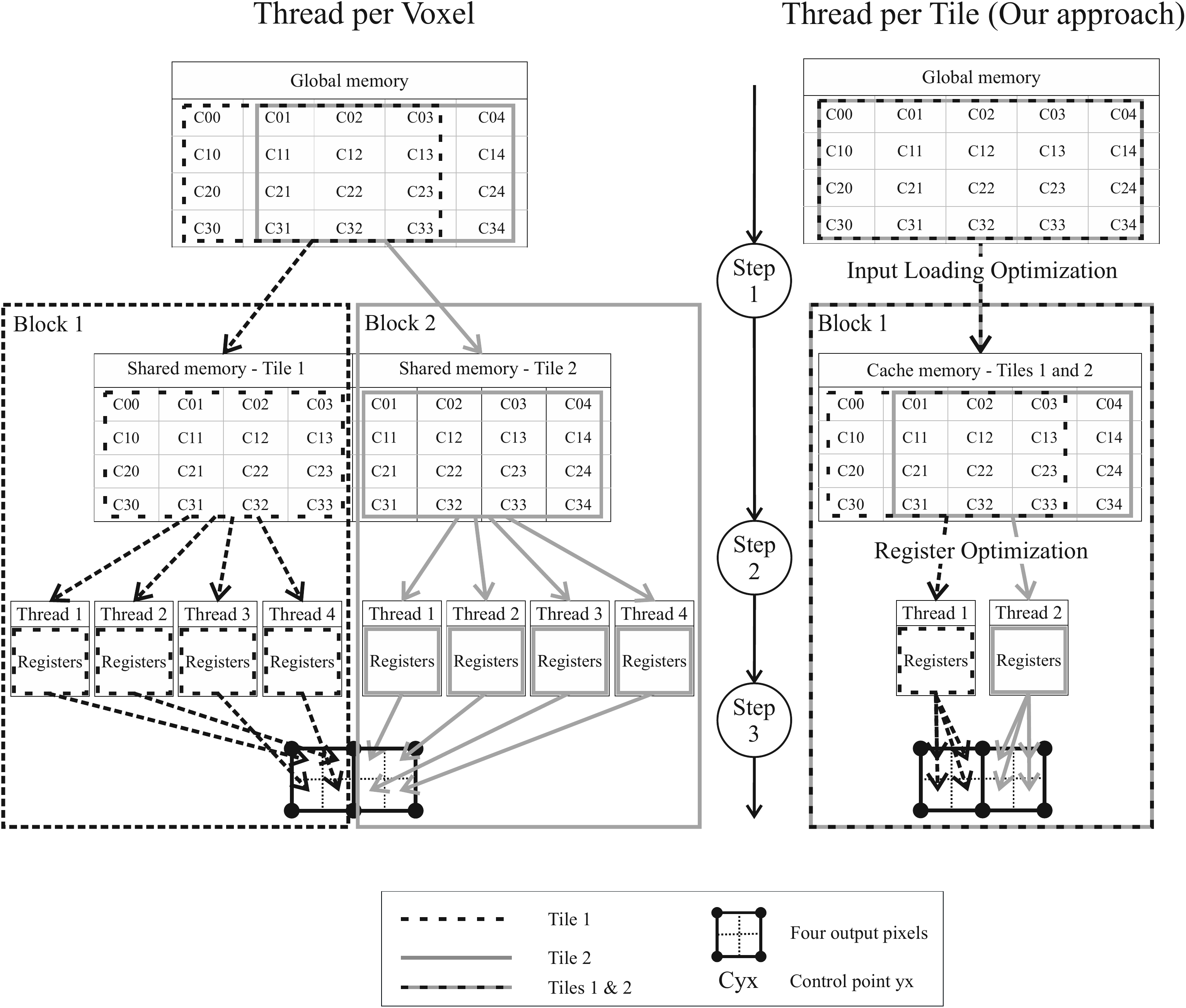}
		%		\caption{Comparison of input loading and register optimization for Thread per Voxel (left) and Thread per Tile (right) for two neighboring tiles. \jgl{Why Thread 1 and Thread 2 appear four times in the left figure?}}
		\caption{Comparison of input loading and register optimization for Thread per Voxel with tiling (left) and Thread per Tile (right) for two neighboring tiles.}
		\label{fig:TT}
	\end{figure*}

	The main idea is to reduce loads from global memory by taking advantage of the overlap of %neighboring tiles.
	tiles assigned to neighboring threads. 
	Figure~\ref{fig:TT} compares the TV approach with tiling (left), explained in Section~\ref{sec:soa}, to our TT approach (right).
	
	In TV, each block of threads works on a unique tile of voxels. Thus, each block requires $ 4^N $ input control points (Section \ref{ssec:tiles}). Therefore, for each tile, we need to move $ 4^N $ control points from global memory to shared memory. 
	\emph{Step 1} in Figure~\ref{fig:TT} (left) illustrates the required data movement from global memory to shared memory for a 2D example. 
	In this example, we have two tiles and each tile is assigned to one block. The two tiles imply the movement of $ 4 \times 4 + 4 \times 4 $ control points from global memory to shared memory.
	
	In TT, we assign one thread per tile to take advantage of overlapping neighboring tiles. %\reduline{we reduce the required amount of transfers from global memory to cache memory}. 
	\emph{Step 1} of Figure~\ref{fig:TT} (right) illustrates the reduction in data movement to \ore{cache} memory, with the overlap in the x-direction. Two tiles require only $ 4 \times 5 $ control points. In 3D medical images, the reduction in data movement is more noticeable, because there is overlap in the three directions. As a result, our approach reduces the data movement from global memory dramatically. \ore{TT requires about 12$\times$ and about 187$\times$ (for  $ 5 \times 5 \times 5 $ tiles) fewer memory transfers in comparison to TV and TH (\ref{sec:mem_transf})}.
	
	\subsubsection{Register optimization} \label{sec:reg_optimiz}
	{The second optimization technique that we apply to TT is based on two main ideas:} 
	%The two main ideas are: 
	1) we load the control points for all voxels of the tile from \ore{cache} memory only once, 
	and 2) we keep the loaded control points in registers, which are the fastest on-chip memory, until thread execution finishes.
	
	%TV
	In TV, threads belonging to the same block work on individual voxels of the same tile. For every voxel belonging to the tile, the corresponding threads need to access %the exactly same area of shared memory as the other threads of the block, in order to load the same set of control points.
	{exactly the same control points as all other threads of the block}. 
	\emph{Step 2} of Figure \ref{fig:TT} (left) illustrates the required data movement from shared memory to registers for a 2D example. In this example, %each tile comprises four pixels and each pixel is assigned to one thread. The figure shows that four pixels require four transfers (from shared memory to registers) of sixteen control points each.
	{each pixel is assigned to one thread, and for every four pixels the corresponding four threads need to read (from shared memory to registers) sixteen control points each (i.e., $4 \times 16$ reads for every four pixels)}.
	
	%TT
	%Shared memory is faster than global memory, but before the GPU can execute any arithmetic operation on the control points, each thread has to move the entire set of control points to its registers first \cite{ProgrGuide,Volkov2010}. To minimize transfers from shared memory to registers, we assign one thread for each tile. 
	{In TT, the one-thread, one-tile assignment minimizes the data movement between \ore{cache} memory and registers}. 
	For all voxels belonging to the tile, the corresponding thread needs to access %the same shared memory area exactly once, in order to load the control points. 
	{from \ore{cache} memory a unique set of control points that is different from the set accessed by any other thread of the block (there is overlap, though).} 
	By utilizing register tiling, the thread keeps the control points in registers, which are faster than \ore{cache} memory~\cite{Volkov2010}, to process every voxel in the tile. \emph{Step 2} of Figure \ref{fig:TT} (right) illustrates the reduction in data movement. %A tile of four pixels requires only one transfer of sixteen control points.
	{For every four pixels, the corresponding thread needs to read only sixteen control points (i.e., $1 \times 16$ reads for every four pixels)}.
	
	\subsection{Thread per Tile with Linear Interpolations (TTLI)} \label{ssec:ttli}
	We extend TT by reformulating the triple sum of Equation (\ref{eq:sum}) to \emph{trilinear} interpolations. The basic idea is that a linear interpolation can replace an addition of two weighted addends. We can extend this to three dimensions, where we combine eight addends into a trilinear interpolation \cite{sigg2005fast}.
	
	We calculate a trilinear interpolation as a combination of seven linear interpolations ({in our implementation,} we do not use the hardware interpolation unit \ore{as this would prevent us from increasing input data locality and output data accuracy (Section \ref{ssec:TH})}).
	\ore{The linear interpolations are beneficial to the performance of our approach because the compiler maps linear interpolations to FMA instructions.}
%	The GPU executes linear interpolations with FMA instructions.
\ore{FMA instructions are preferable for two reasons. First, FMA is more accurate because it executes multiplication and addition in the same step, with a single rounding. Second, FMA is faster because it executes both multiplication and addition with a single instruction \cite{Whitehead2011}.}
	
	Figure \ref{fig:lerp} illustrates the $ 4 \times 4 \times 4 $ neighborhood of control points {that affect a tile of voxels}. %each voxel requires. 
	Each one of the $ 2 \times 2 \times 2 $ colored sub-cubes of control points corresponds to one trilinear interpolation. 
	For each voxel in the tile, the respective thread calculates each one of the eight trilinear interpolations. %corresponding to each one of the colored cubes. 
	The arithmetic operations that are needed for each trilinear interpolation (i.e., colored sub-cube) are independent, %therefore we use 
	{thus enabling} Instruction Level Parallelism (ILP)~\cite{Volkov2010}.
	
\ore{\subsection{Implementation details of TT and TTLI} \label{sec:impl_details}
	Register tiling, which we employ in our approach, requires a careful management of the registers. We explain some of our implementation decisions in the following paragraphs.
	
%	\paragraph{Skipping shared memory}
%	%($< 5\%$) speedup. (Shared memory doesn't play a significant role in our implementation, most of the time is in the loop after loading to registers)
%	According to Perrot et al. \cite{Perrot2016}, at least for the case that the input is overlapping, it may be beneficial to load to registers directly, instead of first loading to shared memory and then from shared memory to registers. The control points needed by each group of tiles can be stored to registers directly (using instead only the hardware-managed caches). With this approach, we avoid pointer arithmetics and synchronization that the management of shared memory requires.	
	
	\paragraph{Register allocation}
	The deformation field of a 3D image requires 64 control points and each control point comprises three values, one for each of the three coordinates (x, y, z). Therefore, we need  $ 3 \cdot 64 = 192 $ registers for the control points only. The control point grid is aligned to the voxel grid and uniformly spaced, therefore we store the scalar B-spline coefficients in Look-Up-Tables (LUTs). TT requires 235 registers in total, whereas TTLI requires 255 registers.
	
	\paragraph{Thread block configuration}
	The amount of required registers limits the maximum active threads per  Symmetric Multiprocessor (SM) to 256 \cite{ProgrGuide}.
	We arrange threads to blocks of $ 4 \times 4 \times 4 $ threads. We select this arrangement because a cube is the geometrical structure that maximizes overlap and consequently minimizes memory transfers (i.e., minimizes Equation \ref{eq:mem_tt} in \ref{sec:mem_transf}).

	\paragraph{Performance at low occupancy}
	Shared and cache memories are slower than registers, therefore TT keeps the control points in registers permanently. We arrange input data in such a way that there are no spills (although in TTLI we have to store a few control points into shared memory). Due to the large amount of registers our approach requires, the occupancy of the GPU falls to 12.5\% for CUDA Compute Capabilities (CC) before 7.x and to 25\% for newer CC \cite{ProgrGuide}. Despite the low occupancy, we can maximize resource utilization by using ILP and avoiding the use of cache memories. Our approach uses a register-only approach to increase the performance substantially~\cite{Volkov2010}.}
	
	%we can't control assembler , at least for  pascal and by official tools
	
	%\subsection{Application of our methodology on CPUs} \label{ssec:cpu}
	\subsection{Application of our approach to CPUs} \label{ssec:cpu}
	%	\jgl{Are VT and VV the same as the TT approach proposed for GPU? You should clarify why the same method is applicable to GPU and CPU. You should clarify if VT and VV can also be called TT.}
	
	{We can apply our TTLI approach to the CPU implementation of BSI. Table \ref{tab:cpu_diff} summarizes the main differences with the GPU implementations.} %\reduline{The techniques that we use for the GPU implementation of BSI can be also applied to the CPU implementations.}
	{Some optimizations are not fully applicable to the CPU implementation, because they %were developed primarily for GPU
		{are tailored to the GPU architecture}. GPUs allow for more fine-grained parallelism in comparison to CPU, which makes GPUs more efficient with small 3D groups of tiles with regards to cache and register management.}
	{We develop two parallel implementations of BSI on CPUs, which take advantage of the several cores and the SIMD units (SSE/AVX) that CPUs have~\cite{agnerCPU, intelIntrinsics}.} 
	%CPUs have SIMD units (SSE/AVX) that execute a Single Instruction on Multiple input Data. 
	SIMD units pack many single values, {which we call \emph{elements},} in a special register, called a \emph{vector}, thus applying a form of register tiling. 
	%Each core of a multicore CPU contains one SIMD unit, thus we use multi-threading to process a significant amount of data simultaneously \cite{agnerCPU, intelIntrinsics}. We create two implementations.
	
	% \usepackage{booktabs}
	\begin{table}[h!]
		\caption{Differences between GPU and CPU implementations {(\checkmark means that an optimization technique is used in the CPU implementation).}}
		\centering
		\resizebox{\columnwidth}{!}{
		\begin{tabular}{lcc}
			\toprule
			Optimization    &                  VT &                  VV \\ \midrule
			Input overlap   & Only in x-direction & Only in x-direction \\
			Register tiling &             Partially &          \checkmark \\
			Linear interpolation  &          \checkmark &          \checkmark \\ \bottomrule
		\end{tabular}
	}
		\label{tab:cpu_diff}
	\end{table}
	
	\paragraph{Vector per Tile (VT)} \label{ssec:vt} %Element per Voxel, Vector per Tile, Element per X direction Voxel, EX, coarse grained
	In this method, we parallelize by using SIMD vectors to simultaneously process many voxels of a tile. {Each thread processes $\delta_{x}$ voxels simultaneously. We iterate through the y,z-dimensions of the tile, $\delta_{x}$ voxels at a time.}
	%	\jgl{How is $\delta_{x}$ related to the SIMD vector length? Do you choose $\delta_{x}$ to be equal to the SIMD vector length?}
	{The drawback of this method is that a SIMD vector is not fully utilized if $\delta_{x}$, a user configurable parameter, is not a multiple of the SIMD vector length.}
	%\jgl{This explanation is too short. No way of understanding how VT is implemented.}
	%\jgl{You have two explain the reasons for the differences with the GPU implementation.} \ope{see beginning of this section}
	
	\paragraph{Vector per Voxel (VV)} \label{ssec:vv} %Element per sub-Cube, Vector per Cube, Vector per Voxel, Element per Cube Vertex, ECV, fine-grained
	In this method, we parallelize by using SIMD vectors to simultaneously process each of the trilinear interpolations a single voxel requires. {This means that, using the SIMD unit,} {each thread processes simultaneously all colored sub-cubes a voxel requires (Figure \ref{fig:lerp}).}
	% 	\jgl{Is the SIMD vector length equal to the number of sub-cubes?}
	{Conveniently, the SIMD vector length is equal to the number of sub-cubes.}
	%\jgl{Explain better and state the differences with VT.}
	%\jgl{Also, you have two explain the reasons for the differences with the GPU implementation.}

	\section{Pre-clinical dataset acquisition} \label{sec:dataset}
	
	%In order to test our approach, we performed a pre-clinical study to solve a pre-clinical application scenario. 
	%Andrea - Corrections look good to me! I would use Juan's rewrite.
	{In order to test our implementations of BSI in a pre-clinical application scenario, we perform a pre-clinical study where we use FFD}.
	%Testing was performed throughout a dataset with 
	{\ore{We create a} dataset \ore{(publicly available)} \cite{dataset} \ore{which} consists of} 
	two sets of subjects and imaging modalities: 1) a patient-specific liver phantom~\cite{Pacioni2015} with DynaCT scanning, 
	and 2) a porcine model with MRI scanning, to validate the registration process in-vivo. 
	Table \ref{tab:data} lists the characteristics of the collected dataset.
	
	{In this section, we describe the dataset in detail. We present evaluation results in Sections~\ref{sec:reg_eval} and~\ref{ssec:qualitative}}.
	
	\begin{table}[h]
		\caption{Image characteristics.} \label{tab:data}
		\centering
		\resizebox{\columnwidth}{!}{
		\begin{tabular}{lccc}
			\toprule
			\makecell[l]{Registration \\ pair} & Resolution & \makecell[r]{Voxel count \\ (millions)} & Voxel Spacing 	 \\ \midrule
			Phantom1          &            512$\times$228$\times$385 &            44.94 &               0.49$\times$0.49$\times$0.49 \\
			Phantom2          &            294$\times$130$\times$208 &             7.95 &               0.90$\times$0.90$\times$0.90 \\
			Phantom3          &            294$\times$130$\times$208 &             7.95 &               0.90$\times$0.90$\times$0.90 \\
			Porcine1          &            303$\times$167$\times$212 &            10.73 &               0.94$\times$0.94$\times$1.00 \\
			Porcine2          &            267$\times$169$\times$237 &            10.70 &               0.94$\times$0.94$\times$1.00 \\ \bottomrule
		\end{tabular}
	}
	\end{table}
	
\begin{figure}[h]
		\begin{minipage}[t]{0.49\linewidth}
			\centering
			\includegraphics[width= \linewidth]{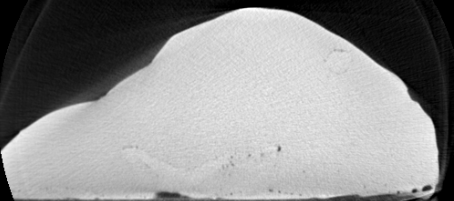}
			(a)
		\end{minipage}
		\begin{minipage}[t]{0.49\linewidth}
			\centering
			\includegraphics[width= \linewidth]{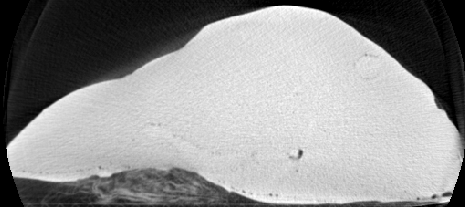}
			(b)
		\end{minipage}\\\\
		\begin{minipage}[t]{0.49\linewidth}
			\centering
			\includegraphics[width= \linewidth]{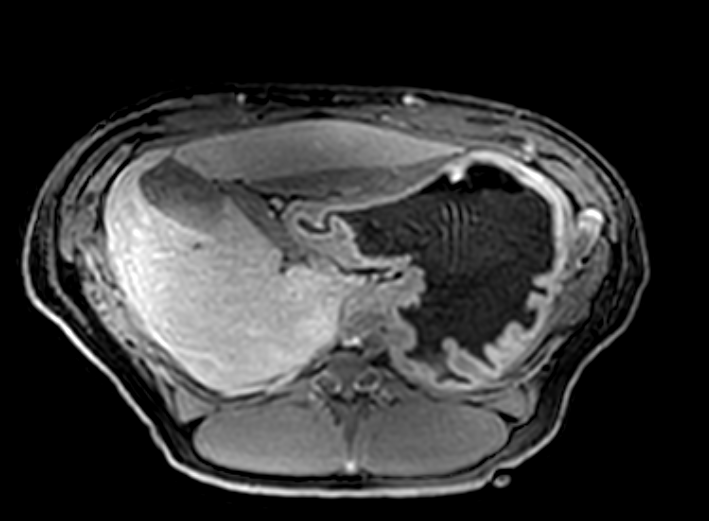}
			(c)
		\end{minipage}
		\begin{minipage}[t]{0.49\linewidth}
			\centering
			\includegraphics[width= \linewidth]{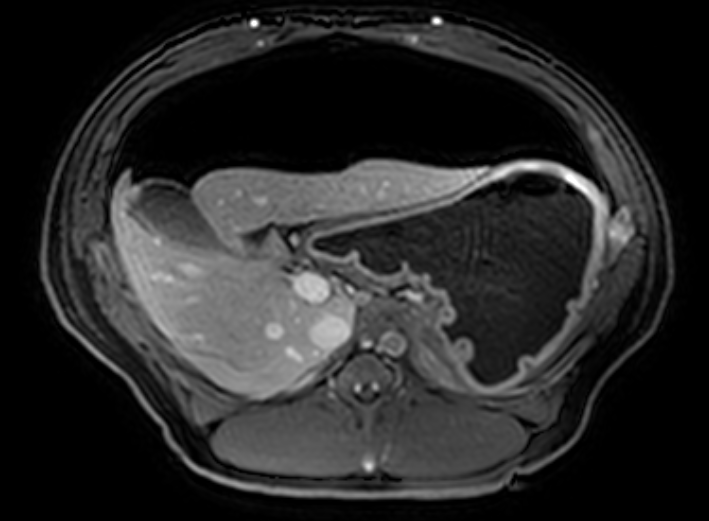}
			(d)
		\end{minipage}
		\caption{Medical images used for pre-clinical evaluation of our optimized image registration through FFD. (a) and (b) show two DynaCT scans of the liver phantom, and (c) and (d) are MRI scans of the porcine model, respectively without (c) and with pneumoperitoneum applied (d).}
		\label{fig:liver_dataset}
	\end{figure}

	\paragraph{Patient-specific phantom of liver}\label{sec:liver_phantom}
	The patient-specific liver phantom presents a total of five tumors and a blood vessel tree. %We use these structures to evaluate the registration process (Section~\ref{ssec:qualitative}). 
	The liver phantom used in our experiments was produced by the ARTORG centre and Cascination\textsuperscript{\circledR}~\cite{Pacioni2015} and has been used by Teatini et al. for registration studies~\cite{andrea2018validation}. We performed three intra-operative CT scans (Artis Zeego, Siemens\textsuperscript{\circledR}) (DynaCT) of the liver phantom in the OR. For each scan, we apply non-rigid deformations to the phantom, which we try to correct through FFD (\emph{Phantom 1, Phantom 2, Phantom 3}). An example of the liver phantom scans is visible in Figure~\ref{fig:liver_dataset}a and Figure~\ref{fig:liver_dataset}b.
	%\jgl{Revise this sentence. It looks incomplete.}. 
	%We use three liver phantom scans 
	%\juan{We use three scans of the liver phantom} (\emph{Phantom 1, Phantom 2, Phantom 3}) with various deformations applied to the phantom.
	
	\paragraph{Porcine model}\label{ssec:pork}
	We performed a porcine study to acquire pre-operative (without pneumoperitoneum) and intra-operative (post pneumoperitoneum) MRI scans. These were used to study the deformation that the liver undergoes due to pneumoperitoneum alone. We performed this study at Oslo University Hospital through the use of a 3T Siemens MRI scanner, model Ingenia Philips\textsuperscript{~\circledR}~\cite{ingenia_MRI}.
	%\jgl{Please add citations for the equipment used. Citations are more important than \circledR}. 
	We performed pneumoperitoneum at 14~mmHg. Both MRI scans were performed with injection of contrast, as done in patients, to improve imaging of the liver parenchyma and blood vessels (Flow rate 5.0 and Volume 11.0, based on the weight of the animal at 55kg). The MRI scans are thin sliced (1.5~mm in \emph{Porcine 1} and 1~mm in \emph{Porcine 2}) enhanced-T1 high-resolution isotropic volume examination (e-THRIVE) scans. The deformation of the liver due to pneumoperitoneum is visible in the differences between images (c) and (d) in Figure~\ref{fig:liver_dataset}\ore{ and further explored in \cite{teatini2019effect}}.
	
	%\jgl{In the caption of the figure: Add a label to each of the four images (a), b), c), d)) and explain them in the text. Right now, Figure 4 is not mentioned anywhere.}
	
	\section{B-spline interpolation evaluation}
	
	{In this section, we evaluate our BSI implementations on GPUs and CPUs in terms of performance and accuracy, and compare them to state-of-the-art implementations.}
	
	\subsection{Evaluation methodology}
	
	\paragraph{Configuration} {In our evaluation}, we use one CPU and two GPUs. The CPU is {a quad-core %(with eight threads) 
		Intel i7-7700HQ@2.8 GHz with HyperThreading. We use gcc v5.4 compiler.}
	%	\jgl{Say number of cores and frequency. Say what compiler.}.
	To show the performance and stability among different GPU generations, we use two GPUs of different generations: 1) NVIDIA GeForce GTX 1050 (with Pascal architecture \cite{ProgrGuide}), and 2) NVIDIA GeForce RTX 2070 (with Turing architecture \cite{nvidiaTuring2018}). We use CUDA SDK v9.2 for the first GPU and v10.1 for the second GPU. %\jgl{Two different SDKs for no reason. Weird, but ok.}. \ope{I would prefer to leave it as is. I use 9.2 on one machine because some tricks wrt registers spills don't work in 10.x (although register spills don't affect performance that much - maybe 5\%). 10.1 in the other machine because I can't use turing with 9.2} 
	We use CUDA event API to acquire the timing results.
	
	\paragraph{Comparison baseline} 
	We compare our approaches to the state-of-the-art BSI {implementations} (Section \ref{sec:soa}). 
	For TH, we use the library from Ruijters et al. \cite{Ruijters2012}. 
	For TV, we create an implementation that %uses tile properties and 
	is based on the recent literature
	%	\jgl{Include citations}
	{\cite{Ellingwood2016,Modat2010,Shackleford2010}}.
	{This implementation of TV uses tiling and is tuned for the GPUs we use.} {We refer to this implementation as \emph{TV-tiling}.}
	%It is a highly optimized version with both literature's observations and our own. 
	%We use 
	{We also compare to} the optimized GPU implementation of the NiftyReg library~\cite{Modat2010}, {which does \emph{not} use tiling}, as GPU reference, and the optimized CPU implementation of NiftyReg~\cite{Modat2010} as CPU reference. {We refer to the NiftyReg implementations as \emph{NiftyReg (TV)}.}
	% 	\jgl{Since NiftyReg is TV without tiling, and your TV uses tiling, I think you should call these implementations "NiftyReg (TV)" and TV-tiling, over the entire section and all graphs.}
	
	{\paragraph{Dataset and metrics}
		We measure the timing information of BSI while applying registration on our dataset.}
	%\reduline{\juan{In our evaluation}} 
	We use two metrics to %evaluate 
	{measure} the performance: 1) \emph{time per voxel} {is} the {execution} time necessary to interpolate a single voxel, and 2) \emph{speedup} %, as 
	{is the performance} improvement over {NiftyReg (TV)}.
	
	\paragraph{Parameters} We select five different tile sizes to evaluate the behavior of the algorithms under different parameters, namely $3 \times 3 \times 3$, $4 \times 4 \times 4$, $5 \times 5 \times 5$, $6 \times 6 \times 6$, $7 \times 7 \times 7$. We select these tile sizes because they are centered around $5 \times 5 \times 5$, which is the default {tile size for} %of 
	non-rigid registration in NiftyReg.

	\subsection{GPU performance}
	
\begin{figure*}[h!]
		\centering
		\begin{subfigure}[b]{.4\textwidth}
			\centering
			\includegraphics[width=\textwidth]{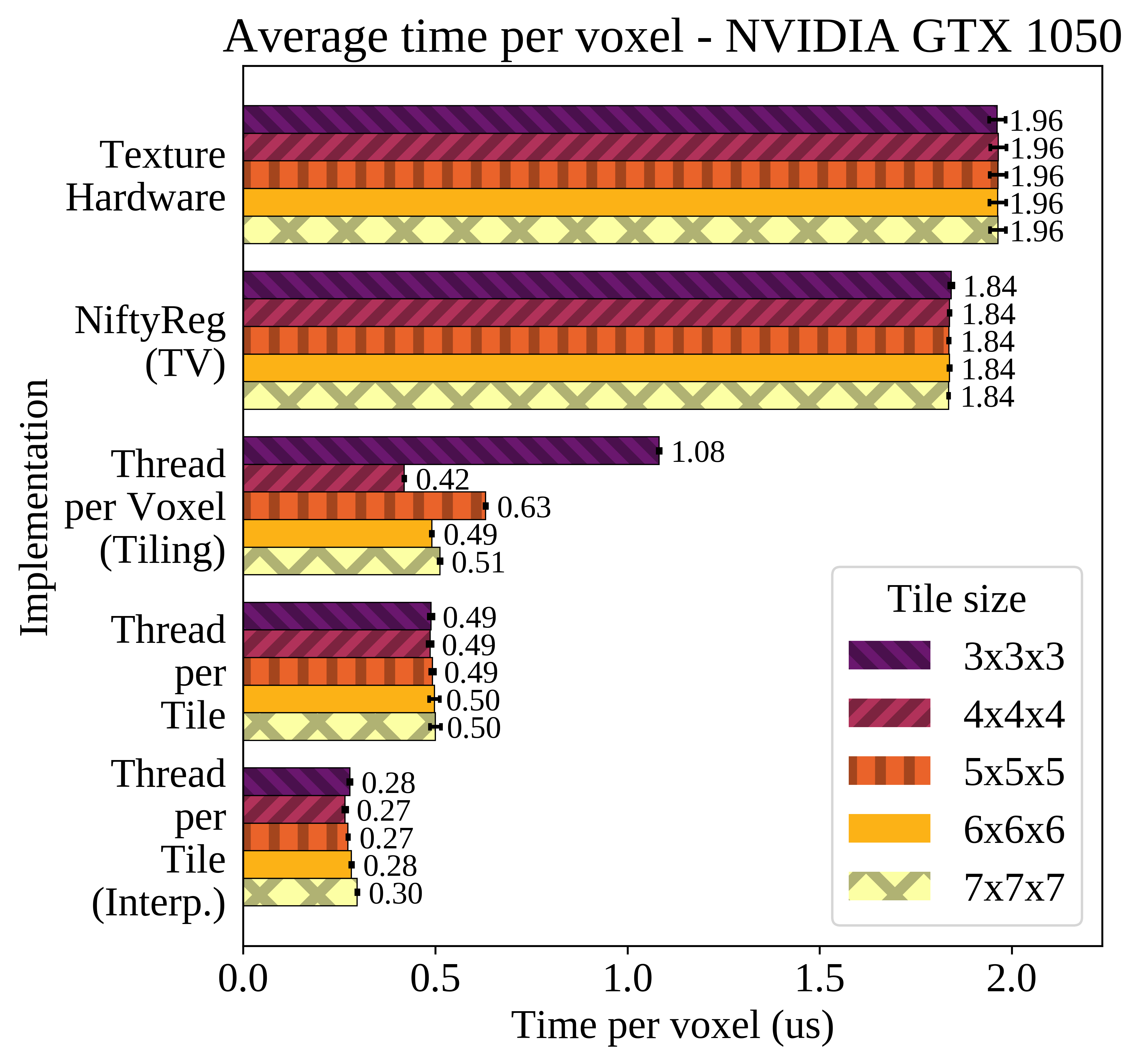}
			\caption{} 
			\label{fig:gpu_times1050}
		\end{subfigure}%
		\begin{subfigure}[b]{.4\textwidth}
			\centering
			\includegraphics[width=\textwidth]{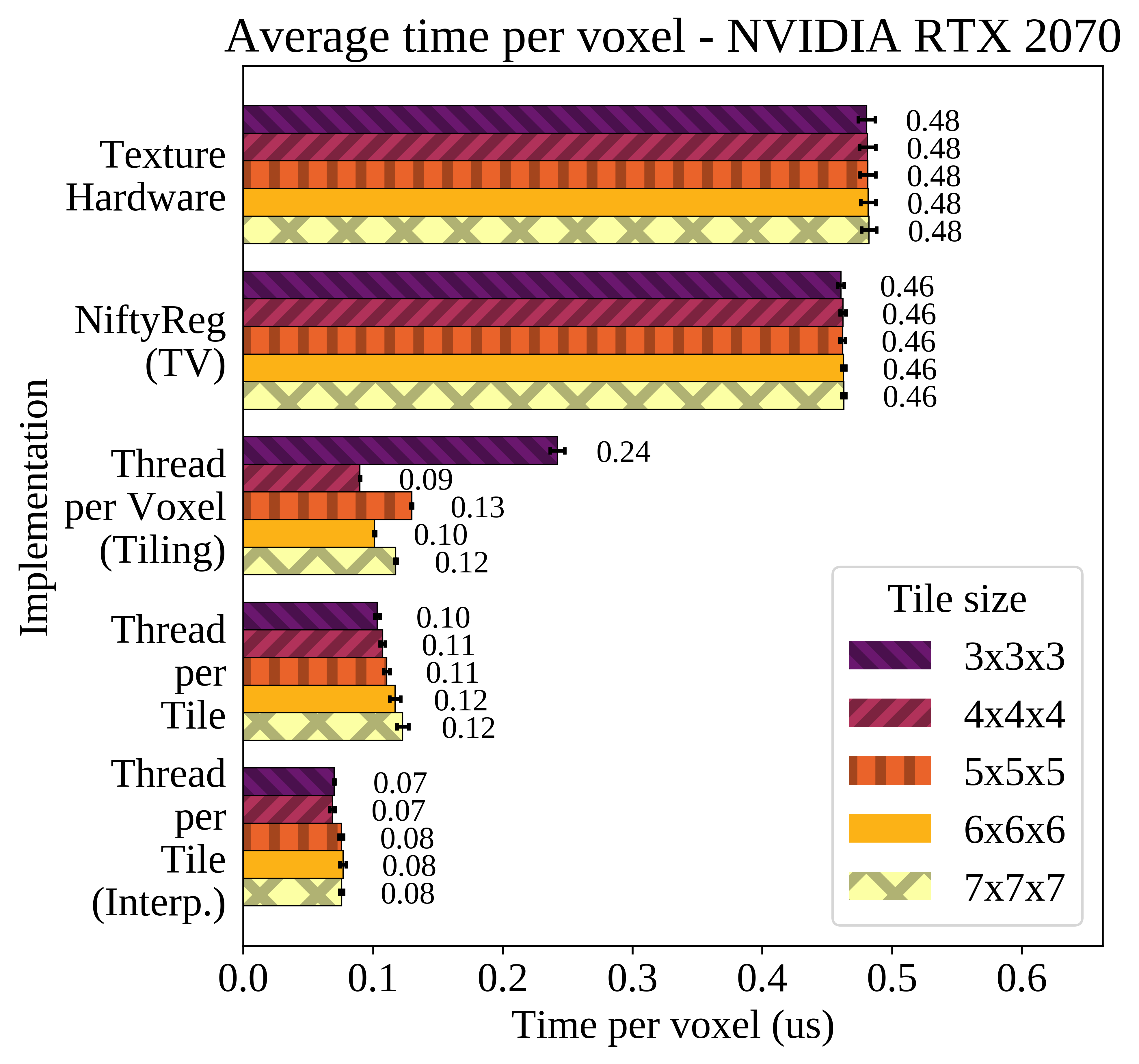}
			\caption{}
			\label{fig:gpu_times2070}
		\end{subfigure}
		\caption{Average time per voxel of the five registration pairs for various tile sizes on GTX 1050 GPU (a) and RTX 2070 GPU (b). {Error bars depict the standard deviation of time per voxel.}
			% 		\jgl{
			% 		These graphs can be more readable: 1) these colors are ugly - try others; 
			% 		2) make label fonts bigger; 
			% 		3) are you showing Min/Max in each bar? If so, say so.
			% 		}
			% 			\jgl{The graphs are disorganized. From top to bottom of the graph put: TH, NiftyReg, TV, TT, TTLI.}
		}  
		\label{fig:gpu_times}
	\end{figure*}

	Figures~\ref{fig:gpu_times1050} and~\ref{fig:gpu_times2070} show the average \emph{time per voxel} for {TH, {NiftyReg (TV)}, {TV-tiling}, TT, and TTLI on the} GTX 1050 and {the} RTX 2070 GPUs, respectively.
	%	\jgl{What are the inputs for these experiments? Are they the dataset in Sect. 4? Clarify.}
	
	% We group times per voxel by implementation and for each group we vary the tile size. With this grouping scheme, we show how each implementation behaves when we vary the tile size. 
	
	%\paragraph{Result analysis for time per voxel}
	%We draw two major conclusions:
	{We make three main observations}.
	{First, TTLI is} {the fastest implementation in all cases}. %\ope{Is this redundant? It is the same observation as in the following paragraph of speedup.}	
	Second, the %performance 
	time per voxel is almost %invariant to 
	independent of the tile size {for all implementations} except {TV-tiling}, for which the thread block size changes with the tile size.
	%	 \jgl{Why is it more sensitive to thread block size? Does this mean that the thread block size changes with the tile size? Clarify}.
\ore{The reasons are three.
	1) Bigger tiles leave more threads inactive at the borders of the image.
	2) Bigger tiles decrease the coalescence of GPU memory accesses. In our approach, a single thread stores an entire tile in the output (Figure \ref{fig:TT}, Step 3).
	3) If the number of SMs does not divide the amount of blocks exactly, some SMs may remain idle (tail effect).
	In conclusion, the performance of our approach in regards to different tile sizes, is a balance between the acceleration that the reduction of data movement offers and the deceleration that border effects and memory uncoalescence cause.}	
	{Third, for all implementations the coefficient of variation (error bars show the standard deviation across the images of our dataset) is less than 3\% which} %shows}
	{reflects} 
	%	\jgl{Are you showing the standard deviation? Where? You mean the Max/Min lines? Btw, "coefficient of variation" is a nice statistic you can mention. Say: "For all implementations the coefficient of variation is less than..."}
	that the image contents do not affect the performance. The reason is that BSI %acts 
	{is regular, i.e., it operates} on all voxels uniformly.
	%	\jgl{No observations about how fast your implementations are? Add them.}
	
\begin{figure*}[h!]
		\centering
		\begin{subfigure}[b]{.4\textwidth}
			\centering
			\includegraphics[width=\textwidth]{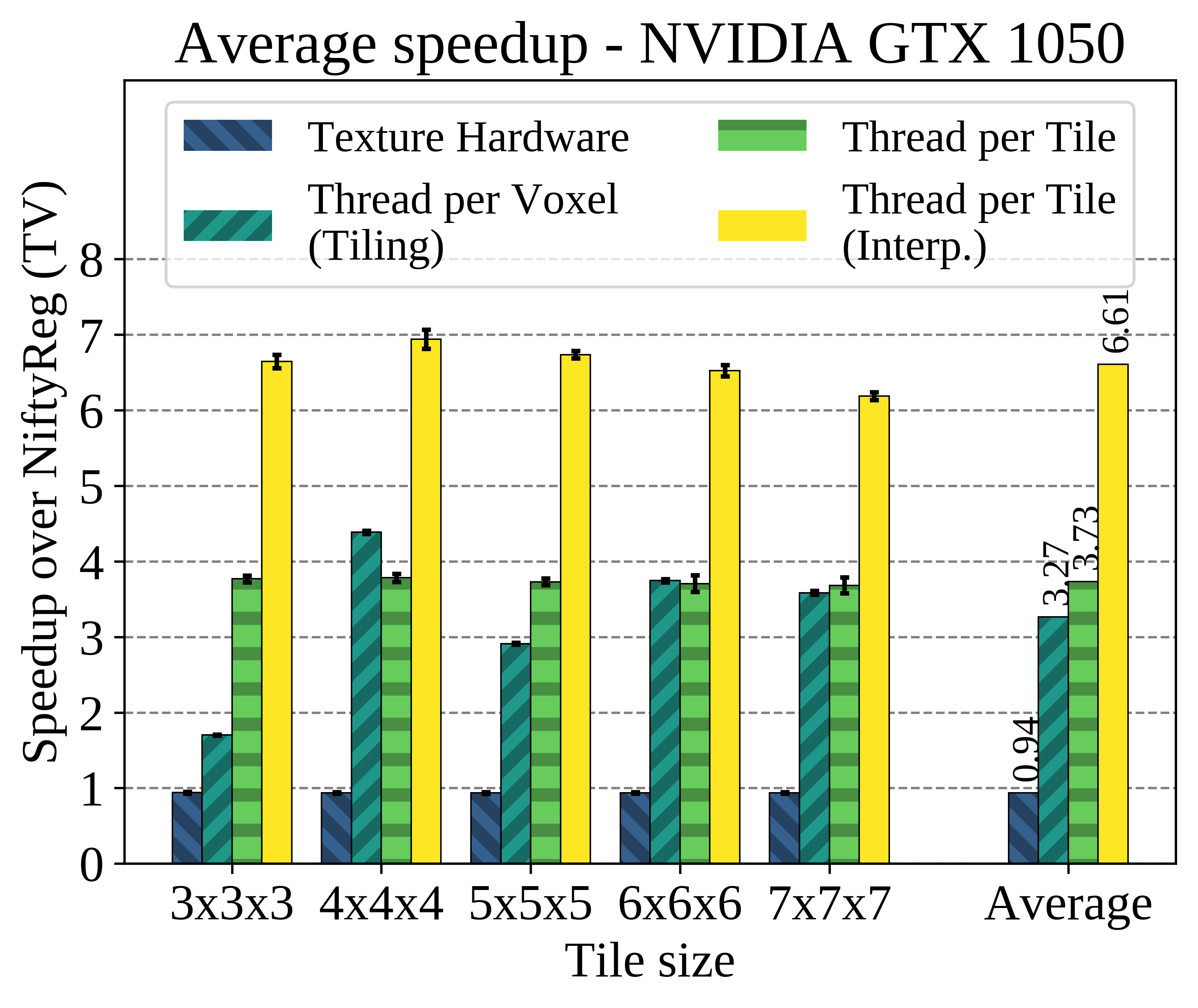}
			\caption{} 
			\label{fig:gpu_speed_big1050}
		\end{subfigure}%
		\begin{subfigure}[b]{.4\textwidth}
			\centering
			\includegraphics[width=\textwidth]{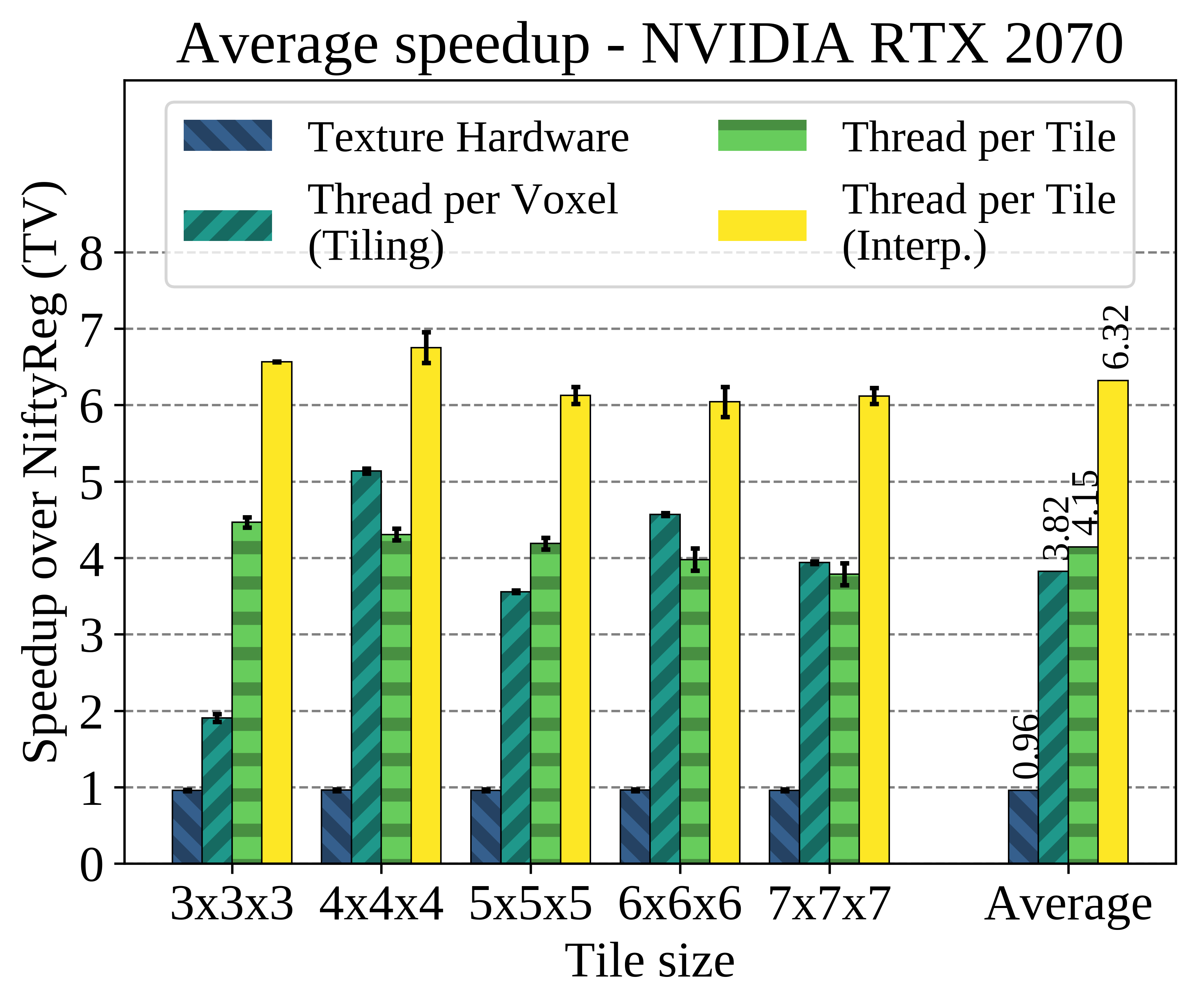}
			\caption{}
			\label{fig:gpu_speed_big2070}
		\end{subfigure}
		\caption{Average %time per voxel 
			{speedup over NiftyReg(TV) for} %of \
			the five registration pairs %for various 
			{with different} tile sizes on {the} GTX 1050 GPU (a) and {the} RTX 2070 GPU (b). {Error bars depict the standard deviation of the speedup.}
			% \jgl{Same comments as for Fig. 5. These colors are better though. However, it is nicer that you have more similar colors for your implementations TT and TTLI, and these are different from TH and TV colors.}
			% 			\jgl{Y-axis label should be "Speedup over NiftyReg(TV)".}
		}
		\label{fig:gpu_speed}
	\end{figure*}

	Figures \ref{fig:gpu_speed_big1050} and \ref{fig:gpu_speed_big2070} show the average \emph{speedup} {over NiftyReg(TV)} for {TH, {TV-tiling}, TT, and TTLI on the} GTX 1050 and {the} RTX 2070 GPUs, respectively. 
	%We show how each implementation behaves in comparison to NiftyReg.
	%We group speedup by tile size and for each group we vary the implementation. With this grouping scheme, we show how each implementation behaves in comparison to NiftyReg. 
	
	%\paragraph{Result analysis for speedup} 
	%We draw three major conclusions:
	{We make \ore{two} observations.} 
	First, our TTLI approach %is outperforming the rest 
	%\juan{outperforms all other implementations} in all cases. %and 
	%\juan{TTLI} 
	{is 6.5$\times$ {(up to 7$\times$)} faster than NiftyReg(TV), on average}. 
	{TTLI outperforms the second fastest (TT) by an average of 1.77$\times$ on GTX 1050 and 1.5$\times$ on RTX 2070.}
	%	\jgl{For what other implementation are those 1.56 and 1.3? Clarify}.
	\ore{Second}, {TTLI} %works well 
	{shows similar speedups over NiftyReg(TV)} on both Pascal architecture (GTX 1050) and Turing architecture (RTX 2070) GPUs, {which demonstrates that our optimizations are widely applicable and performance-portable}.

\ore{\subsubsection{Analysis of performance limitations}
	%TODO take care of 2nd observation of speedup.
This section describes the limitations that define the performance of our approach.

TT does not %improve a lot 
provide significant speedup over TV-tiling. The reason is that our TT approach reduces data movement significantly, which
makes TT compute-bound.
We observe with the NVIDIA's Visual Profiler~\cite{ProfGuide} that the compute utilization of TT is at about 90\% of the peak.
Since the amount of computation in TT is not reduced with respect to TV-tiling, the potential improvement is limited.

Reformulating the summation of TT to trilinear interpolations (Section \ref{ssec:ttli}) reduces the computational complexity of Equation (\ref{eq:sum}) to half (\ref{sec:computational}) and increases the usage of FMA instructions. TTLI is 50\% - 80\% faster than TT.
After removing the computational intensity problem, TTLI is no %more 
longer compute-bound. The main bottleneck is the uncoalescence of the output (Figure \ref{fig:TT}, Step 3). In our experiments, fixing the uncoalescence proved more computationally costly than the uncoalescence itself.
% \jgl{What does "latency" mean here? You mean this is the only remaining bottleneck? If so, explain why you can't solve this uncoalescence.}

Thread divergence, caused by the inactive threads at the borders of the image, reduces the computation throughput for both TT and TTLI.

With $5 \times 5 \times 5$ tile, TTLI achieves 670 GFLOP/s and 62 GB/s on {the} GTX 1050~\footnote{\ore{NVIDIA profiler (version 2019.4.0) does not provide metrics for counting FLOPs on {the} RTX 2070.}}. %(NVIDIA does not provide the necessary tools for counting FLOPs on RTX 2070 currently). 
The empirical limits \cite{yang2018empirical} of {the} GTX 1050 are 2091 GFLOP/s and 95 GB/s. We observe that TTLI is close to the bandwidth limit, but not so close to the computation limit.

%TODO is this paragraph necessary or did I just include it for the reviewer?
% With $5 \times 5 \times 5$ tile, TT achieves 887 GFLOP/s with 35 GB/s memory bandwidth and TTLI 670 GFLOP/s and 62 GB/s on {the} GTX 1050~\footnote{\ore{NVIDIA profiler (version 2019.4.0) does not provide metrics for counting FLOPs on {the} RTX 2070.}}. %(NVIDIA does not provide the necessary tools for counting FLOPs on RTX 2070 currently). 
% The empirical limits \cite{yang2018empirical} of {the} GTX 1050 are 2091 GFLOP/s (or 1045 GFLOP/s without FMA instructions) and 95 GB/s. We make two observations. First, TT is close to the computational limit. Second, TTLI 
% \jgl{Why do you include these limits if you don't comment on them?}
}
	
	\subsection{CPU performance} \label{ssec:speedup_cpu}
\begin{figure*}[h!]
		\centering
		\begin{subfigure}[b]{.4\textwidth}
			\centering
			\includegraphics[width=\textwidth]{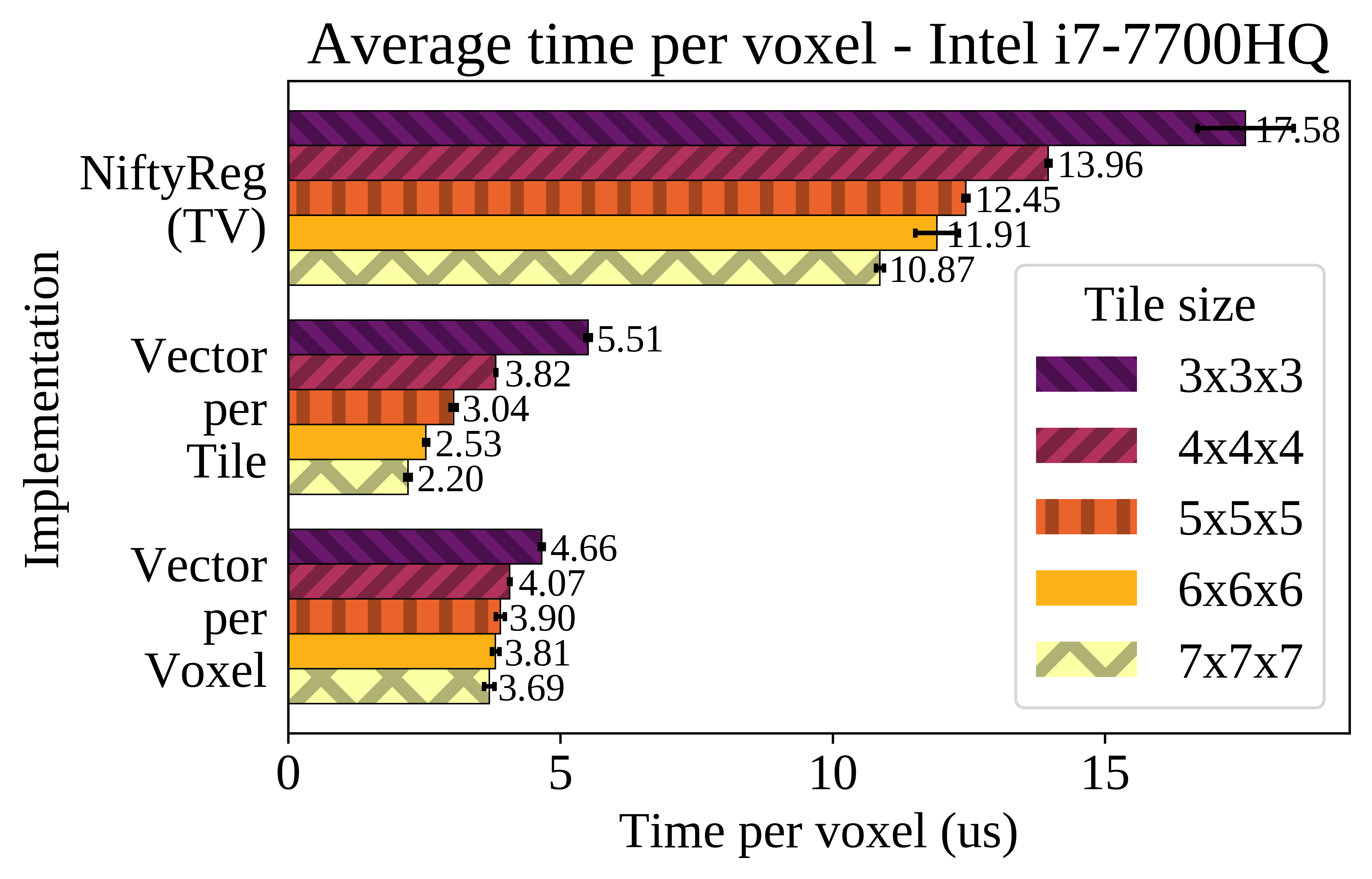}
			\caption{} 
			\label{fig:cpu_times7700}
		\end{subfigure}%
		\hspace{0.1cm}
		\begin{subfigure}[b]{.37\textwidth}
			\centering
			\includegraphics[width=\textwidth]{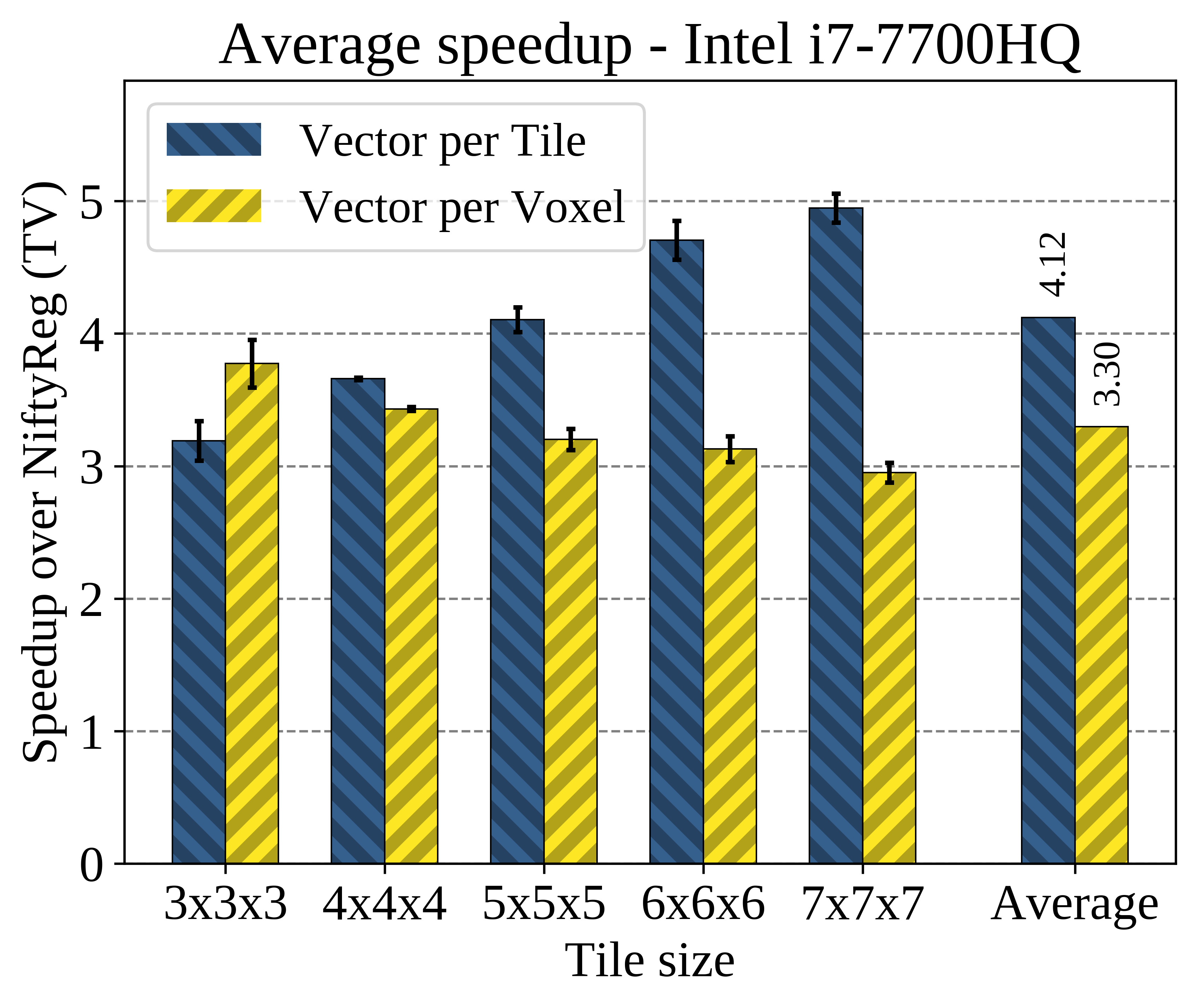} 
			\caption{}
			\label{fig:cpu_speed_big7700}
		\end{subfigure}
		\caption{Average time per voxel (a) and speedup (b) of BSI for various tile sizes using our implementation of BSI on CPUs. Error bars depict the standard deviation.}  
		\label{fig:cpu_times_speed}
	\end{figure*}

	We apply %the methodology of our GPU approach to CPU 
	{our approach to BSI to our CPU implementations} (Section \ref{ssec:cpu}). 
	%Figures \ref{fig:cpu_times7700} and \ref{fig:cpu_speed_big7700} show the mean and standard deviation of time per voxel and speedup of the CPU approaches. We group speedup by tile size and for each group we vary the implementation.
	{Figures~\ref{fig:cpu_times7700} and~\ref{fig:cpu_speed_big7700} show respectively time per voxel and speedup results of our CPU approaches for different tile sizes.}
	
	%\paragraph{Result analysis}
	%We compared the performance of our CPU implementations in terms of \emph{time per voxel} and \emph{speedup}. Our CPU implementations outperform the reference in all cases by 3$\times$ to 5$\times$. We conclude that our methodology can be beneficial even for CPU.
	We make \ore{four} observations. 
		First, our CPU implementations (VT and VV) outperform the baseline NiftyReg (TV) by an average of 4.12$\times$ and 3.30$\times$, respectively.
%		\ore{Second, VT generally outperforms VV. An exception is the $3 \times 3 \times 3$ tile. In this case, the small $\delta_{x}$ leaves a large part of the SIMD vectors unutilized (Section \ref{ssec:vt}).}
		Second, for all implementations, larger tiles result in lower time per voxel, as they can take more advantage of the CPU cache hierarchy. This effect is more pronounced in VT, which achieves a speedup of almost 5$\times$ for the largest tiles.
		\ore{Third, the speedup of VT increases as the tile size increases because bigger tiles fill more \emph{slots} of the SIMD vectors. VT is the fastest option when more than 3 slots are filled.
		Fourth, the speedup of VV does not increase, as the time per voxel of NiftyReg decreases with faster rate than the time per voxel of VV. VV is the recommended option only for $3 \times 3 \times 3$ tiles.}

%	\subsubsection{\ore{Effects of varying tile sizes}}
%%TODO take care of 2nd observation (replace?). VT achieves better speedup because of more slots used. Or integrate the followint to the analysis.	I can tell how the other methods are affected by the tile size (get them from extended paper)
%\ore{We make two observations in regards to the effect of tile size to performance.
%	
%	First, in VV, execution time decreases with increasing tile size. The reason is that the bigger the tile is, the more the voxels that use the same control points are (Section \ref{ssec:vv}). Thus, we have fewer memory transfers with bigger tiles.
%	
%	Second, the performance of VT improves with increasing tile size for two reasons. 1) Bigger tiles fill more slots of the SIMD vectors (Section \ref{ssec:vt}) and therefore each CPU thread calculates more voxels simultaneously. 2) Bigger tiles require fewer transfers from memory (\ref{sec:mem_transf}).
%
%	In conclusion, larger tiles result in lower time per voxel, as they can take more advantage of the CPU cache hierarchy.}

	\subsection{Accuracy}
	%To show the accuracy benefits of utilizing FMA instruction in our linear interpolation approaches, we compare each implementation with a high precision CPU implementation. Table \ref{tab:accuracyGPU} shows the average absolute difference between each \emph{GPU} approach and the high precision CPU implementation. Similarly, Table \ref{tab:accuracyCPU} shows the average absolute difference between each \emph{CPU} approach and the high precision CPU implementation.
	{Our implementations employ FMA instructions, which are more accurate than regular multiplications~\cite{ProgrGuide}, in the calculation of linear interpolations. 
		In this section, we show the accuracy improvements that stem from FMA instructions. 
		We {create} a high precision CPU implementation {by using double precision arithmetic (64-bits floating point numbers) and we use this implementation as} reference.} 
	%	\jgl{Explain where this high precision CPU implementation is coming from. Why is it high precision? And include citation/s.}
	
	{Tables~\ref{tab:accuracyGPU} and~\ref{tab:accuracyCPU} show respectively the average absolute error of all GPU implementations and all CPU implementations with respect to the high precision CPU implementation}. 
	
	\begin{table}[h]
		\caption{Average absolute error of BSI approaches on GPUs with respect to a high precision CPU implementation.}
		%\jgl{Citation needed}
		\label{tab:accuracyGPU}
		\centering
		\begin{tabular}{lr}
			\toprule
			Implementation                & Error ($ e^{-6} $) \\ \midrule
			Texture Hardware         &               9245 \\
			Thread per Voxel (Tiling)&                5.5 \\
			NiftyReg (TV) GPU        &                5.3 \\
			Thread per Tile          &                5.6 \\
			Thread per Tile (Interp.)&                2.8 \\ \bottomrule
		\end{tabular} 
	\end{table}
	
	\begin{table}[h]
		\caption{Average absolute difference of BSI approaches on CPUs with respect to a high precision CPU implementation.}
		%~\cite{}\jgl{Citation needed}.}
		\label{tab:accuracyCPU}
		\centering
		\begin{tabular}{lr}
			\toprule
			Implementation                & Error ($ e^{-6} $) \\ \midrule
			NiftyReg (TV) CPU        &                6.0 \\
			Vector per Tile          &                3.0 \\
			Vector per Voxel         &                3.0 \\ \bottomrule
		\end{tabular} 
	\end{table}
	
	%\paragraph{Result analysis}
	We draw three conclusions.
	First, our implementations that %use 
	{employ} FMA instructions (i.e., TTLI on GPUs, VT and VV on CPUs) are almost two times more accurate than the rest. %implementations that do not use FMA instructions. %The other GPU implementations are slighty more inaccurate because they are implemented as a serial summation.
	Second, TH is significantly less accurate than the rest of the implementations, as expected from the low accuracy of interpolation hardware \cite{ProgrGuide}. {TH is 3300$\times$ less accurate than TTLI.}
	Third, %GPU is competing well with CPU in regards to accuracy.
	{most GPU implementations show accuracy values in the same order of magnitude as CPU implementations.}

	\section{Registration evaluation} \label{sec:reg_eval}
	
	{In this section, we evaluate the performance impact of our BSI implementations on the overall registration process.}
	
	\subsection{Evaluation methodology}
	To test the contribution of our %accelerated B-spline interpolation 
	{BSI implementations} to the total time required for the registration of medical images, we integrate our TTLI approach into NiftyReg\footnote{\url{https://github.com/oresths/niftyreg_bsi}}~\cite{Modat2010}. %(Section \ref{sec:soa})
	\ore{The control points in NiftyReg correspond to a coarse deformation field. We calculate the fine deformation field (i.e., the displacement of all voxels) by interpolating the coarse deformation field using BSI.}
	\ore{We} compare \ore{the total registration time with our BSI} %with 
	{to the original NiftyReg} registration, on {our dataset presented in} %of 
	Section~\ref{sec:dataset}. 
	We %test 
	{evaluate} the performance of non-rigid registration on two platforms: 
	a) a quad-core Intel i7-7700HQ@2.8 GHz CPU (with HyperThreading) and a GTX 1050 GPU, and 
	b) a six-core Intel i7-8700@3.2 GHz CPU (with HyperThreading) and an RTX 2070 GPU.
	%The tile size is 
	{We set the tile size} to $5 \times 5 \times 5$, {which is} the default setting %of 
	in NiftyReg. 
	%\reduline{In order to have a fair comparison between our two platforms, we limit the number of iterations of the cost function of the registration} \jgl{No idea what this sentence means. Clarify. Cite as needed}.\ope{The idea is that registration is an iterative function. The number of iterations varies depending on where the algorithms finds a local optimum. The GPU implementations are different so they end in different local optima, without necessarily one optimum being better than the other. One implementation may terminate "sooner" (in less iterations) if it falls early in an optimum, while the other continues for more iterations (and most likely it has found a better optimum). The implementation that terminates sooner will give better times, although the results probably is worse. So I limit both implementations to the same number of iterations, effectively terminating both "sooner". Now, I had a paragraph explaining all these things but I removed it because it seems to me it's not the most important thing to include in 3500 words. Therefore maybe it's better to remove this comment completely}

	\subsection{Performance evaluation}
\begin{figure}[h!]
		\centering
		\includegraphics[width=0.79\linewidth]{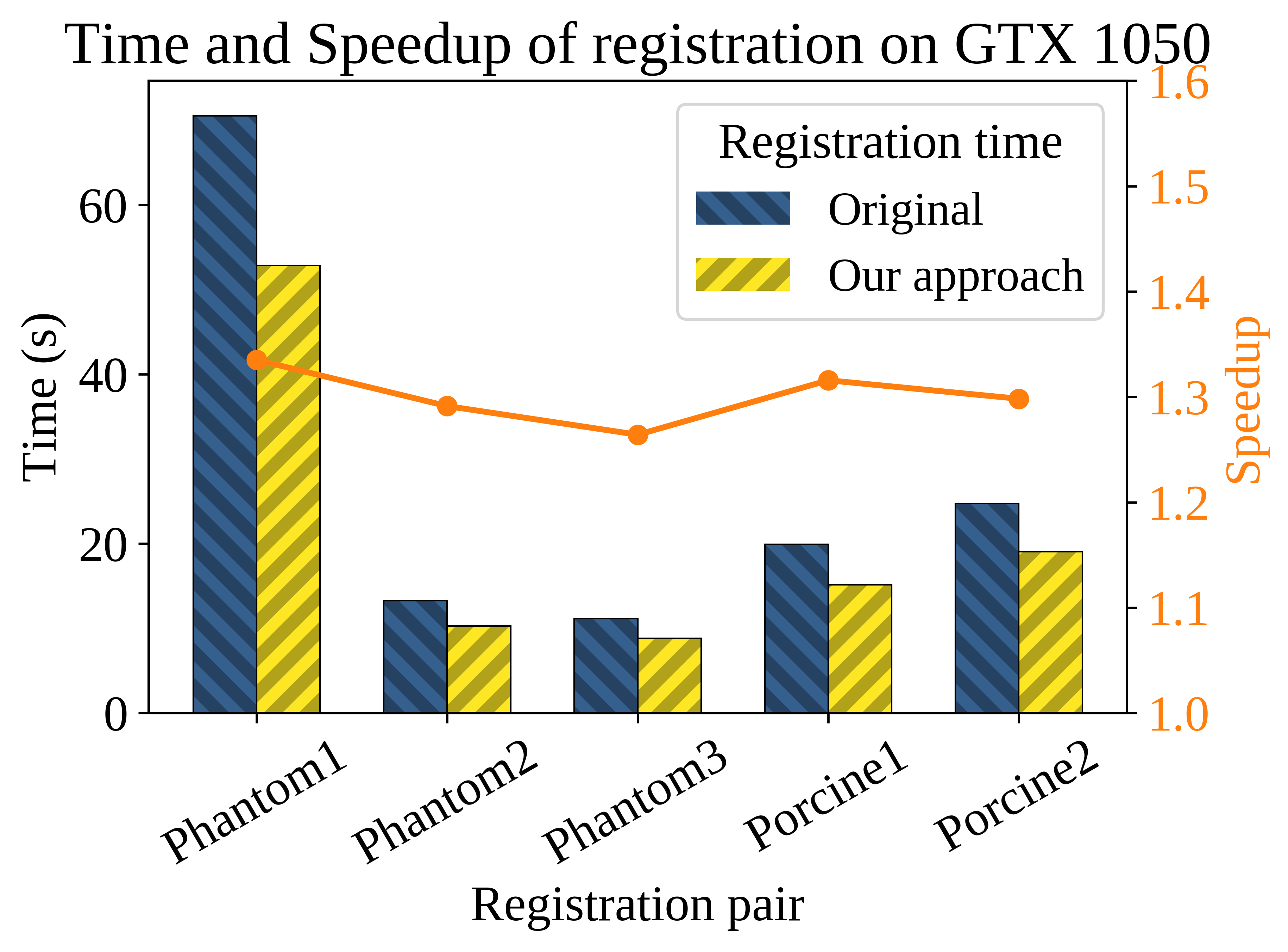}
		\caption{Time and speedup of registration with our improved BSI GPU approach on GTX 1050.}
		\label{fig:time_speed1050}
	\end{figure}

\begin{figure}[h!]
		\centering
		\includegraphics[width=0.79\linewidth]{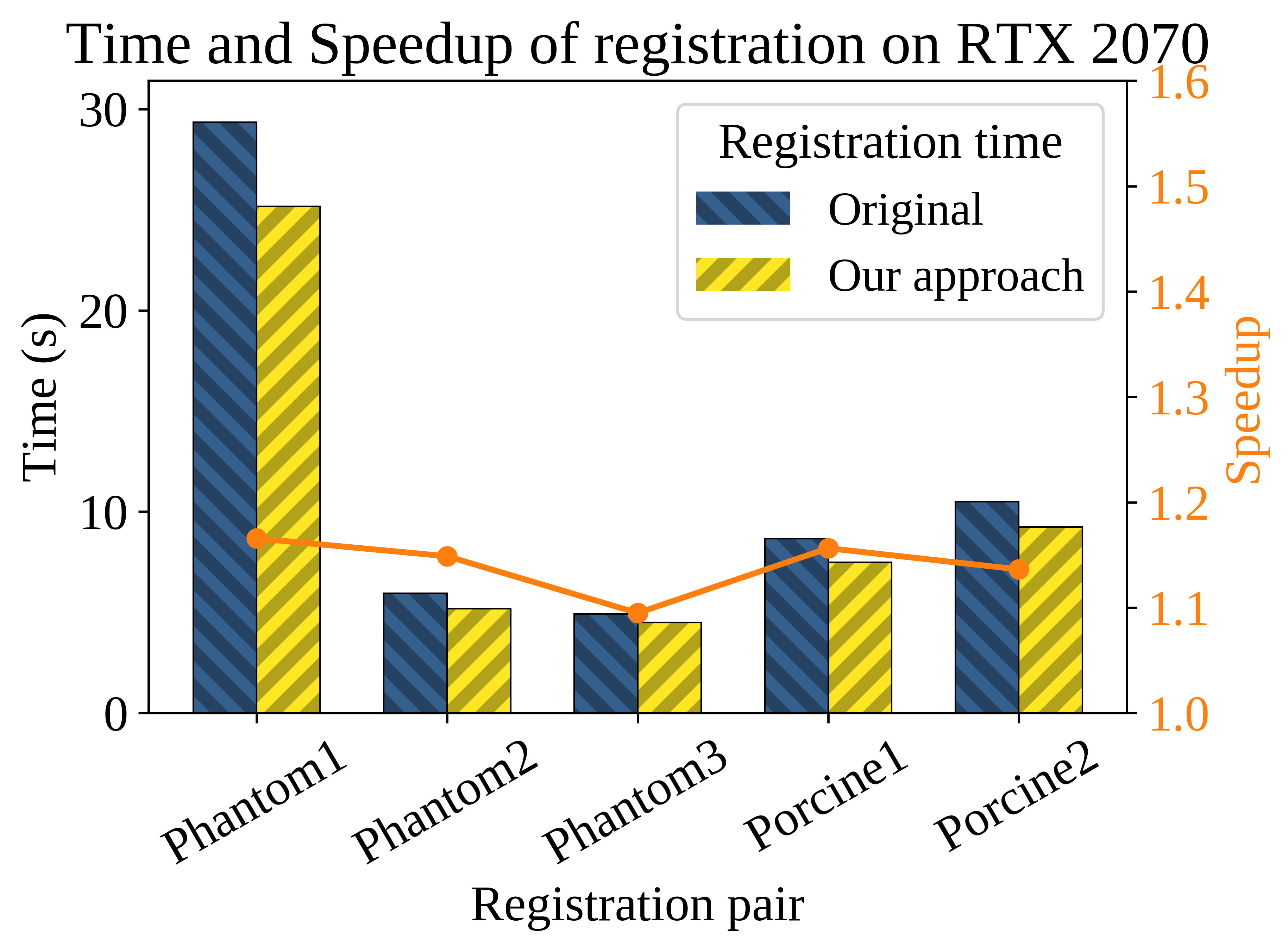}
		\caption{Time and speedup of registration with our improved BSI GPU approach on GTX 1050.}
		\label{fig:time_speed2070}
	\end{figure}

	Figures~\ref{fig:time_speed1050} and~\ref{fig:time_speed2070} show the total registration time and the speedup of our approach on %GTX 1050 and RTX 2070 respectively.
	{the two platforms}.
	%	\jgl{Tables are ugly. Plot two figures: one for GTX1050 and one for RTX2070. In each figure, show bars for the execution times of Original and Our approach (left y-axis), and a line for the speedup (right x-axis).}
	
	We draw two major conclusions. 
	First, registration with our BSI approach is faster in all images %, on both GPUs.
	{on both platforms}.
	The speedup of registration is 1.30$\times$, on average, on {the platform with a} GTX 1050 GPU, %whereas the speedup on RTX 2070 is 1.14$\times$ on average.
	{and 1.14$\times$ on the platform with an RTX 2070 GPU.}
	Second, although the performance %increase 
	{improvement of our BSI approach} is almost the same for both GPUs, we do not observe the same results %with regards to the performance of image registration. 
	{for the entire image registration.} 
	The reason %is that when executing on RTX 2070, BSI represents a smaller time portion of the registration (Amdahl's law \cite{amdahl1967validity}). Therefore, the performance gains will be better visible on a fully GPU-optimized registration workflow.
	{resides in Amdahl's law \cite{amdahl1967validity}: 
		while BSI represents 27\% of the total registration time on the platform with a GTX 1050 GPU, it takes only 15\% on the platform with an RTX 2070 GPU}.% \jgl{Include those percentages.}
	{As a result, the overall performance impact on the registration workflow depends on the characteristics of the compute platform.}

	\section{Clinical validation of image registration}
	
	%\jgl{Include a short paragraph explaining the purpose of this section.}
	%\jgl{Btw, I think this section could be a subsection of Section 6.} \andre{I disagree with this chapter being part of section 6, it's an important structure and part of the evaluation of the registration, especially since we are targeting a journal with medical applications. Instead, I agree with respects to the introductory sentence/paragraph, however I am a bit worried about the word limit of CM (3500 words). We can use some sentences from Section 4: e.g. "As aforementioned, we validated our implementation of accelerated FFD onpre-clinical dataset described in Section 4." }
	In this section, we present the validation of our implementation of accelerated FFD on our pre-clinical dataset described in Section \ref{sec:dataset}.
	
\begin{figure*}[h]
		\centering
		\begin{minipage}[t]{0.49\linewidth}
			\includegraphics[width= \linewidth]{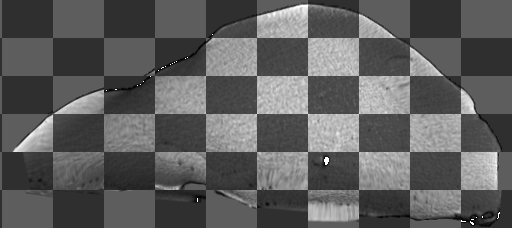}
		\end{minipage}
		\begin{minipage}[t]{0.49\linewidth}
			\includegraphics[width= \linewidth]{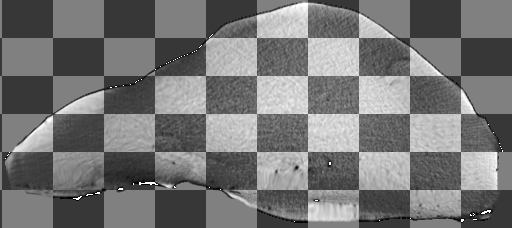}
		\end{minipage}
		\caption{
		Comparison of registration through qualitative checkerboard assessment on liver phantom scans. (Left) shows the registration results using an affine registration. (Right) shows the results of non-rigid FFD using our BSI implementation.
		\label{fig:liver_phantom_checker}}
	\end{figure*}

\begin{figure*}[h]
		\centering
		\begin{minipage}[t]{0.49\linewidth}
			\includegraphics[width=1\linewidth]{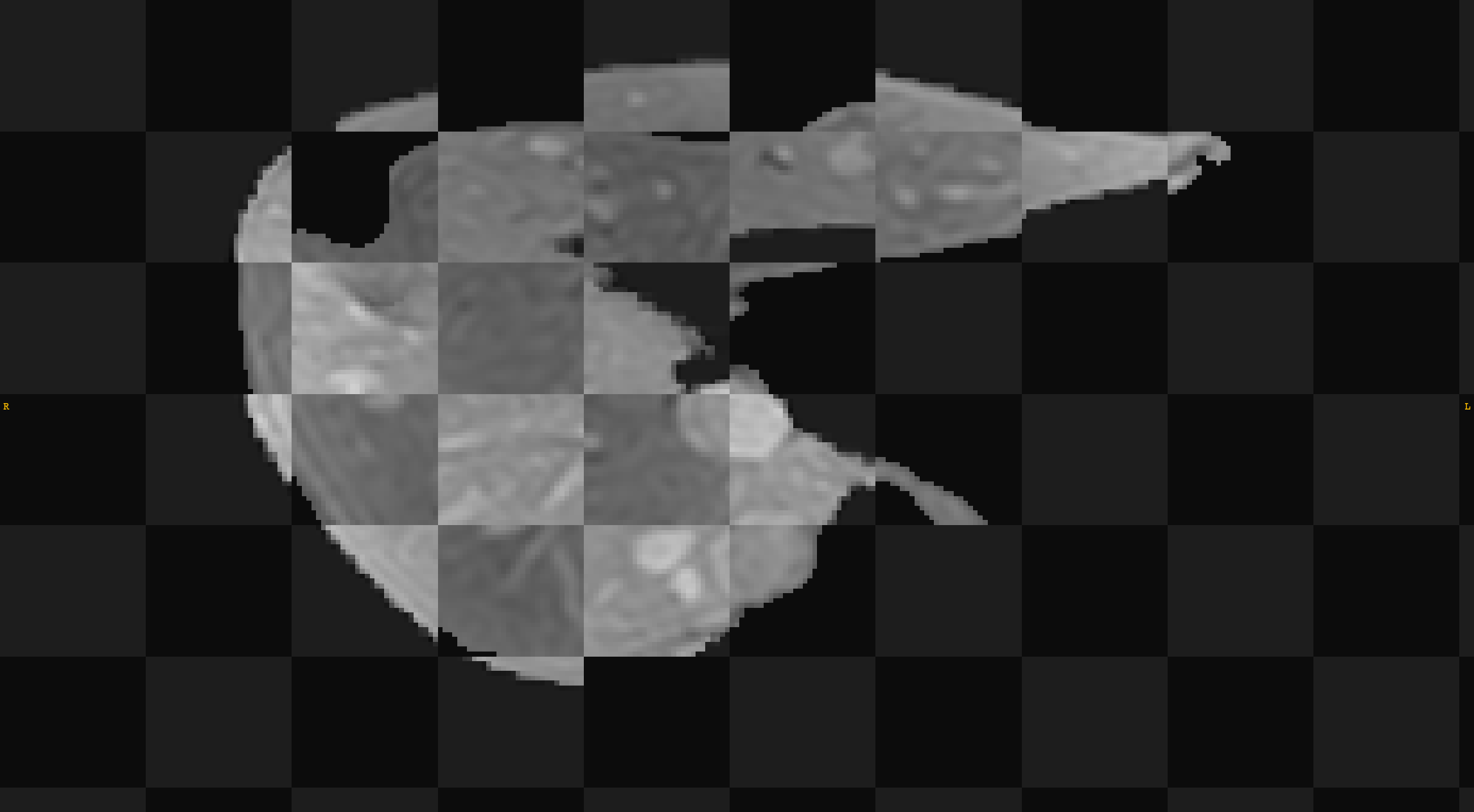}
		\end{minipage}
		\begin{minipage}[t]{0.49\linewidth}
			\includegraphics[width=1\linewidth]{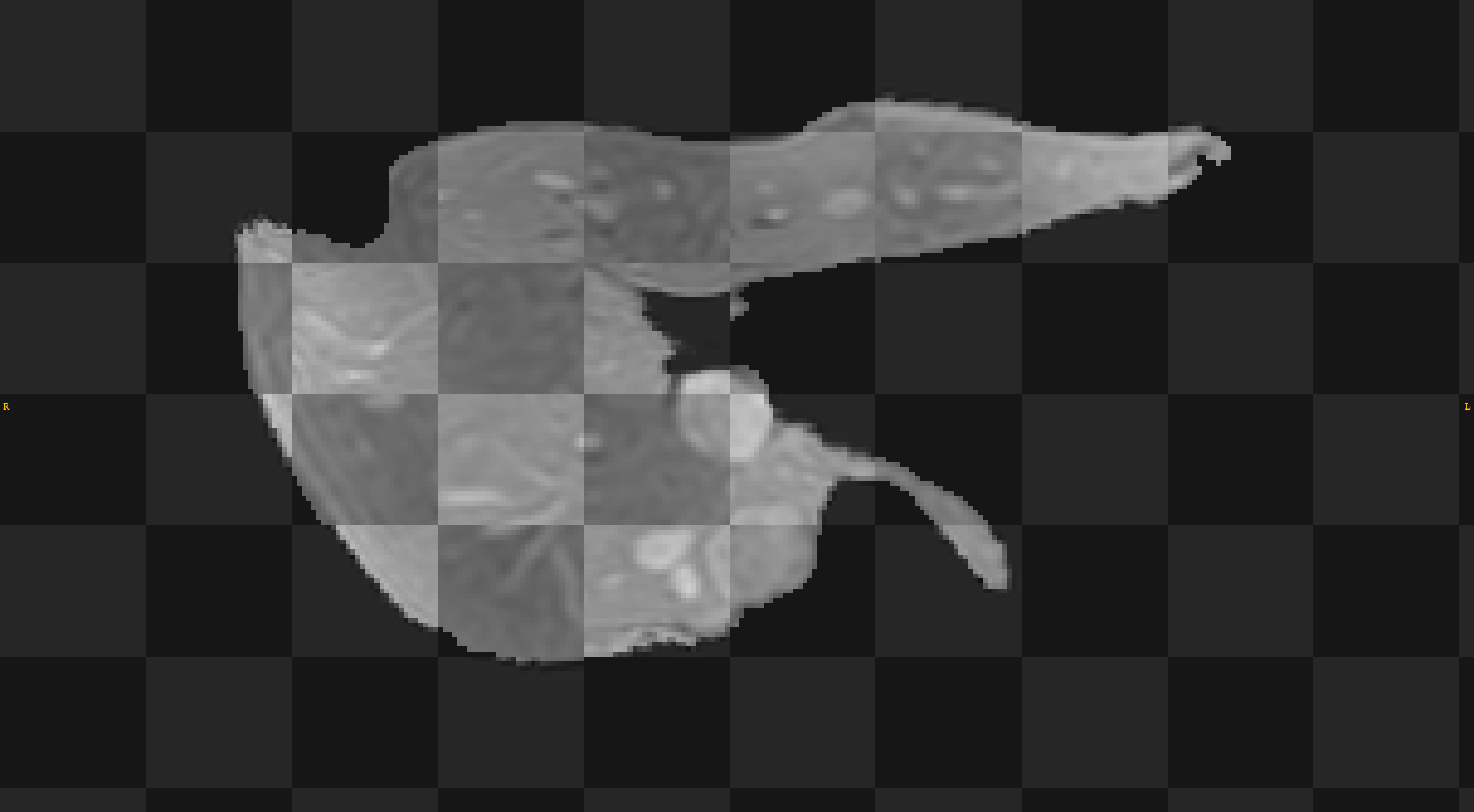}
		\end{minipage}
		\caption{
		Comparison of registration through qualitative checkerboard assessment on porcine liver scans. 
		{(Left) shows the registration results} using an affine registration. 
		{(Right) shows the results of} non-rigid FFD using our BSI implementation. 
		\label{fig:liver_pig_nonr_checker}}
	\end{figure*}

	\paragraph{Qualitative assessment} \label{ssec:qualitative}
% 	\jgl{Need to introduce the difference image figures. Write qualitative comments for difference? (maybe the comments about the checkerboard are sufficient for difference images as well)}
	We perform qualitative assessment of the registration using a checkerboard validation procedure~\cite{pluim2016truth}. Our method provides accurate registration for the parenchyma (the outer shape of the liver is preserved correctly) for both the liver phantom and porcine model. Tumors and vessel structures of the phantom are consistent between images \ore{(Figure \ref{fig:liver_phantom_checker})} and approximately also vessel structures for the porcine model are correctly registered (Figure \ref{fig:liver_pig_nonr_checker}).
	
	\begin{figure*}[h!]
		\begin{minipage}[t]{1\linewidth}
			\includegraphics[width= \linewidth]{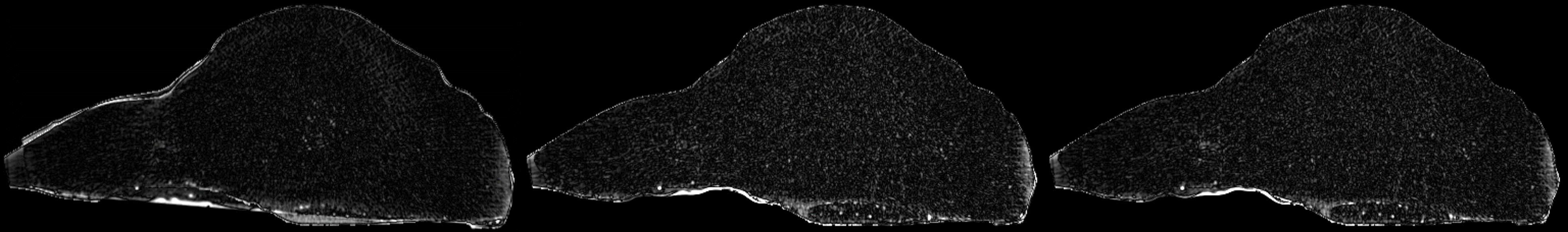}
		\end{minipage}
		\caption{
		Comparison of registration through quantitative difference image assessment on liver phantom scans. (Left) shows results using an affine registration; (Center) shows the results of non-rigid FFD using our BSI implementation; (Right) shows the results of non-rigid FFD using original NiftyReg.
		\label{fig:liver_phantom_diff}}
	\end{figure*}
	
	\begin{figure*}[h!]
		\begin{minipage}[t]{1\linewidth}
			\includegraphics[width= \linewidth]{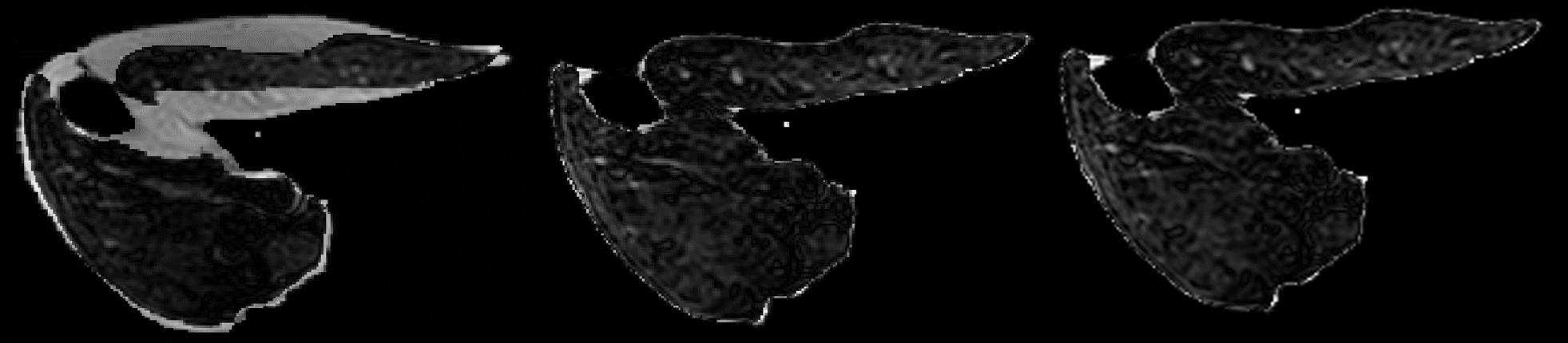}
		\end{minipage}
		\caption{
		Comparison of registration through quantitative difference image assessment on porcine liver scans. (Left) shows results using an affine registration; (Center) shows the results of non-rigid FFD using our BSI implementation; (Right) shows the results of non-rigid FFD using original NiftyReg.
		\label{fig:liver_pig_nonr_diff}}
	\end{figure*}

	\paragraph{Quantitative assessment}
    % Difference images provided a means of comparison between the datasets, in particular, the authors compare the output from the registration using NiftyReg, and our approach; Comparing in both case the output of non-rigid registration with the affine registered image.
%    Images were normalised to account for contrast changes and different image intensities.
    	
%    The difference image outputs between non-rigid with our method and NiftyReg had a mean of 0.019, whereas 0.216 were on average the differences found for the affine compared to non-rigid methods. 
    We create normalized difference images between the output of the registration and the target intra-operative image for three registration approaches: 1) affine, 2) proposed, and 3) original NiftyReg (Figures \ref{fig:liver_phantom_diff} and \ref{fig:liver_pig_nonr_diff}). 
    % As expected, there are large differences when comparing affine to non-rigid implementations and unperceivable differences between the two non-rigid registration approaches.
    Table \ref{tab:TableSSIM} shows the mean absolute error (MAE) for all images of our dataset.
    As expected, the mismatch to the target intra-operative image is greater with affine than with non-rigid registration approaches. The two non-rigid registration approaches perform almost equally (the average MAE across the five image pairs is 0.216 for affine, 0.1240 for our approach and 0.1249 for original NiftyReg.
    
    In order to quantify how the different registration approaches affect the accuracy of the registration as output images, we apply Structured Similarity Index Metric (SSIM) \cite{hore2010image} to our dataset. With the SSIM, we measure the similarity between the output of the registration approach and the target intra-operative image (Table \ref{tab:TableSSIM}).
    
\begin{table}[h!]
\centering
    	\caption{Mean absolute error (Left) on normalised outputs of affine registration and non-rigid registration with our approach and original NiftyReg, using the intra-operative image as reference. Structured Similarity Index Metric (Right) of the registration output, using the intra-operative image as reference).
    	\label{tab:TableSSIM}}
    \resizebox{\columnwidth}{!}{
\begin{tabular}{@{}lcccccc@{}}
\toprule
          \multirow{2}{*}{\makecell[l]{Registration \\ pair}} & \multicolumn{3}{c}{MAE} & \multicolumn{3}{c}{SSIM}     \\ \cmidrule(lr){2-4} \cmidrule(lr){5-7}
          & Affine    & Proposed    & NiftyReg    & Affine & Proposed & NiftyReg \\ \midrule
Phantom 1 & 0.229     & 0.13        & 0.131       & 0.865  & 0.929    & 0.934    \\
Phantom 2 & 0.234     & 0.172       & 0.179       & 0.916  & 0.952    & 0.946    \\
Phantom 3 & 0.256     & 0.174       & 0.172       & 0.889  & 0.952    & 0.95     \\
Porcine 1 & 0.201     & 0.072       & 0.072       & 0.797  & 0.912    & 0.911    \\
Porcine 2 & 0.162     & 0.072       & 0.071       & 0.716  & 0.737    & 0.737    \\ \midrule
Average  &0.2164 & 0.1240 &0.1249 & 0.8368        & 0.8963      & 0.8956        \\\bottomrule
\end{tabular}
}
\end{table}

    We make  three observations. First, the non-rigid registration approaches have much higher similarity than the affine registration approach. Second, our approach and the original NiftyReg have almost equal similarities.
    % (the mean difference between the SSIM across the images is 0.0027, as compared to affine transforms where the mean difference is around 0.06).
    Third, our approach gives slightly better similarity than the original NiftyReg approach.
    % (average SSIM is 0.8963 for our approach and 0.8956 for original NiftyReg).
    Further evaluation of accuracy of the registration can be inferred from the original studies performed by Modat et al. \cite{Modat2010}.
	%\jgl{No quantitative results in a quantitative assessment section?\andre{I normally just take a look at how the checkerboards match in terms of shape of the liver and vessel structures.}}
	
\section{Discussion}
	In this work we optimize BSI and integrate it to FFD to accelerate the performance of medical image registration. However, our improved BSI can also be used in generic image interpolation applications, %like 
	e.g., image zooming~\cite{unser1999splines}, by using image pixels as the control points.
	
	The performance of image registration can be further improved by merging the other steps of FFD with B-spline interpolation.
	%	As an additional benefit of merging other steps, the uncoalescence of the output of BSI could possibly be avoided.
	%TODO I have to mention uncoalescence first
	By optimizing the rest of the registration process, the execution time of the registration further diminishes, enabling new possibilities for fast intra-operative updates without intra-operative CT acquisitions, e.g., through liver models reconstructed with US~\cite{clements2015validation} or through stereo video reconstructions~\cite{andrea2018validation}.
	
	The speedup of image registration through optimized FFD is important not only for pneumoperitoneum compensation, but also for compensation of several other deformations that the liver commonly undergoes during surgery. If real-time registration is possible, FFD can be used in IGS to compensate for deformations that result from lifting the liver with a surgical instrument or resecting liver ligaments (liver mobilization). %TODO Maybe comment about real-time like in email
	
	A limitation of our current implementations is that they work only with control point grids that are aligned to the voxel grid and uniformly spaced. Uniform spacing is usually sufficient for medical images~\cite{Ellingwood2016,Shackleford2010}. Support for non-uniform grids is possible with minimal changes (e.g., calculating B-spline basis functions weights on-the-fly). {We leave this support for future work.}
	
	\section{Conclusion}
	
	%What has been done in the paper to what purpose (for solving what problem) and the key advantages of the proposal
	%In this work, we present a new thread assignment scheme that reduces memory traffic significantly compared to other state-of-the-art B-spline interpolation approaches.
	{This paper presents our approach to B-spline interpolation, which is optimized to reduce data movement.}
	%We achieve it by assigning one GPU thread per tile. This assignment has two key advantages. 
	{The key idea of our approach is to assign one worker thread per tile of voxels.
		This has two main advantages. }
	First, %the loading of the input demonstrates significant overlap. 
	{data movement during input loading is significantly reduced.} 
	Second, %the control points, that each group of tiles needs, stay permanently in registers. 
	{the input control points can be kept in registers during the entire computation.} 
	To further enhance the performance of our implementation, we rearrange the weighted summation of control points into trilinear interpolations. %This rearrangement has two key advantages. 
	{This results in two key advantages.} 
	First, %it reduces 
	{the trilinear interpolations reduce} the computational load. 
	Second, %it increases 
	{they increase the} {interpolation} accuracy. 
	%We apply our optimized approach to non-rigid registration of medical images to enhance the performance of registration in time critical applications, like IGS.
	
	%The key results
	%The results confirm that by restructuring the BSI algorithm to reduce the number of memory transfers is indeed beneficial to the performance. In addition, the use of trilinear interpolation proves helpful not only for performance, but also for the total accuracy. 
	{Our experimental} evaluation on two sets of subjects and imaging modalities shows that our BSI approach offers improved performance and accuracy with respect to state-of-the-art implementations.
	%The proposed approach, 
	TTLI, {our best approach on GPUs}, %\reduline{improves accuracy by up to $ 3300 \times $ and }
	performs up to $ 7 \times $ faster in comparison to the other GPU implementations. {Our implementations that use trilinear interpolations perform approximately 2$\times$ better than the other in regard to {interpolation} accuracy.}
	%	\jgl{Where are these two numbers coming from? I don't see them in previous sections.}
	
	% 	the improved performance reduces the computation time of BSI and therefore speeds tasks such as: volume rendering, volume reconstruction and image registration. The clinical workflow may benefit from the reduced computational time provided by our implementation by faster visualization of 3D models after segmentation
	
	We integrate our BSI approach into {the} NiftyReg medical image registration library and validate it in a pre-clinical application scenario. 
	Our approach improves the performance of non-rigid {image} registration by 30\% and 14\%, {on average}, on our two platforms with a GTX 1050 GPU and an RTX 2070 GPU, respectively. The improved performance reduces the computation time of image registration. Therefore faster updates of the organ and its structures are possible during IGS.
	
	% Updates of the resection lines can also be modified intra-operatively based on the newly registered models.
	
	%The final conclusion (basically stating that your proposal is good for that problem)
	%We conclude that our methodology improves the performance of BSI, in terms of both speed and accuracy. Our approach is an efficient replacement of BSI in non-rigid registration. 
	{As a result, non-rigid registration of medical images can benefit from our BSI approach on GPUs to greatly enhance the performance and accuracy of registration in time-critical applications (e.g., image guided surgery)}.
	
			% 	use section* for acknowledgement 
	\section*{Acknowledgment}
	This work is supported by High Performance Soft-tissue Navigation ( HIPERNAV - H2020-MSCA-ITN-2016 ). HIPERNAV has received funding from the European Union’s Horizon 2020 research and innovation programme under grant agreement No 722068. The authors would also like to thank the radiology staff at the Intervention Centre, Oslo University Hospital, who collaborated to perform the animal experiment on the porcine model. Juan G\'omez-Luna and professor Onur Mutlu would like to thank VMware and all other
industrial partners (especially Facebook, Google, Huawei, Intel, Microsoft) of the SAFARI research group.
	
	% or
	
	
	%\appendices
	%% you can choose not to have a title for an appendix
	%% if you want by leaving the argument blank
	%\section{}
	
\ore{	\appendix
	\section{Off-chip memory to on-chip memory data movement}  \label{sec:mem_transf}
	
%	\jgl{Shouldn't the appendices go after the references? Please check in other papers of this journal. ore:After acknowledge, before references.}
	
	We use the external memory model \cite{Kim2012} to describe the data movement from off-chip memory to on-chip memory. We consider a 3D image. Let us define $ M $ as the total number of voxels, $ N=64 $ as the number of control points, $ T $ as the number of voxels inside each tile, and $ L $ as the size, in words (words are 32-bits long, a common size for storing integer and real numbers), of transactions into the cache (i.e., transactions between off- and on- chip memory).
	%Note that if $ L $ is larger than four control points the rest is wasted. That is because the location of the data in the memory usually will be too far apart since every four control points the y or z axis changes (we have a $ 4 \times 4 \times 4 $ cubic neighborhood). On the contrary, if the CUDA kernel is organized in blocks of tiles, $ L $ can be as large as $ 4+block\_dimension-1 $.
	The $ L $ sized memory transfers of the three cases we are interested in are:
	
	a) \emph{No tiles}: When we do not have tiles, for each of the M voxels, we need to transfer N control points from global memory to shared memory. Transfers happen in L sized chunks. Hence, the total number of transfers required is
	\begin{equation} \label{eq:mem_noTiles}
	\frac{N \times M}{L}
	\end{equation}
	
	b) \emph{Hardware trilinear interpolation}: 
	Each voxel is affected by the $4^3$ control points surrounding it. However, if we use the texture unit to get their trilinear interpolations directly, only $2^3$ loads are required \cite{sigg2005fast}.
	Therefore, when we utilize the texture hardware for loading the input, for each of the M voxels, we need to transfer $ 2^3 $ control points from global memory to cache memory. Transfers happen in L sized chunks. Hence, the total number of transfers required is
	\begin{equation} \label{eq:mem_th}
	\frac{2^3 \times M}{L}
	\end{equation}
	
	c) \emph{A block per tile}: When we use a block for each tile, for each tile we need to transfer N control points from global memory to shared memory. Each tile contains T voxels, thus the total number of tiles is $ M/T $. Transfers happen in L sized chunks. Hence, the total number of transfers required is
	\begin{equation} \label{eq:mem_tv}
	\frac{N \times M}{T \times L}
	\end{equation}
	
	d) \emph{Blocks of tiles}:
	When we have 3D blocks of tiles, and each block contains $ l \times m \times n $ tiles, for each block we need to transfer $ (4+l-1) \times (4+m-1) \times (4+n-1) $ (Section \ref{sec:input_optimiz}) control points from global memory to shared memory (or cache). Each block contains $ l \times m \times n $ tiles and each tile contains T voxels, thus the total number of blocks is $ M/( l \times m \times n \times T ) $. Transfers happen in L sized chunks. Hence, the total number of transfers required is
	\begin{equation} \label{eq:mem_tt}
	\frac{(4+l-1) \times (4+m-1) \times (4+n-1) \times M}{l \times m \times n \times T \times L}
	\end{equation}

	\paragraph{Observations}
	We make the following four observations. First, a hardware trilinear interpolation implementation requires fewer memory transfers than a no tiles implementation because $ 2^3 < N $ in all cases. Second, a block per tile implementation requires fewer memory transfers than a hardware trilinear interpolation implementation because $ N/T < 2^3 $ when $T > 8$. $T > 8$ is a rare case (T is 125 by default in NiftyReg). Third, a blocks of tiles implementation requires fewer memory transfers than a block per tile implementation because $ \frac{(4+l-1) \times (4+m-1) \times (4+n-1)}{l \times m \times n} < N $ as long as a block contains more than one tile. Fourth, the CPU implementations are a special case of Equation (\ref{eq:mem_tt}), in which $l=m=1$, i.e., each thread processes contiguous tiles in the x-axis direction.

	\section{Computational complexity} \label{sec:computational}
	In order to evaluate the arithmetic performance of TTLI and TT, we perform the computational analysis of both implementations in this section.
	
	\paragraph{TT}
	For every voxel of the output image, we need to calculate the triple sum in Equation (\ref{eq:sum}). Each operand of the summation requires the multiplication of one control point ($\phi$) with three weights ($B$). Thus, each voxel requires $ (64\ summands)*(3\ multiplications + 1\ accumulation)-1=255 $ vector ($\phi$ is a  3D vector in deformation fields) arithmetic operations. The calculation of Equation \ref{eq:sum} requires $4+4+4=12$ scalar loads for the weights and 64 vector loads for the control points. If we use one weight for the $B_l(u) \cdot B_m(v) \cdot B_n(w)$ product, instead of three individual weights, the required operations decrease to $(64\ summands)*(1\ multiplications + 1\ accumulation)-1=127$ (same as a parallel reduction) and the weights to be loaded increase to $4*4*4=64$. This is not suitable for our register-only implementations, because there are not enough registers to store the 64 weights and the use of one of the caches would impact the performance substantially (Section \ref{sec:impl_details}).
	
	\paragraph{TTLI}
	For every voxel of the output image, we reformulate the summation of the $ 4 \times 4 \times 4 $ weighted control points to trilinear interpolations. We divide the $ 4 \times 4 \times 4 $ cubic neighborhood to eight $ 2 \times 2 \times 2 $ sub-cubes, as in Figure \ref{fig:lerp}. Each sub-cube corresponds to a trilinear interpolation. A trilinear interpolation requires seven linear interpolations for its calculation. A linear interpolation has the form $ a+w*(b-a) $, which equals to a subtraction and a fused multiply-accumulate (FMA) operation. Thus, for the eight sub-cubes and the ninth final sub-cube that is formed by the eight results of the eight trilinear interpolations, we have $ (9\ cubes)\times(7\ linear\ interpolations.)\times(2\ operations) = 126 $ operations for each voxel.
	%As for the loads, nine scalar ones for the weights and 64 vector ones for the control points are required.

	\paragraph{Observations}
	Without taking into consideration instruction dual-issue, $\Theta(n)$ equals to 255*(number of voxels) and 126*(number of voxels) respectively.}

	%Weights: Let us suppose the dimension of a tile is $ n_x \times n_y \times n_z $. For linear interpolation 9 different floating point weights are required per voxel and $ 3*(n_x + n_y + n_z) $ for all voxels of the tile. For triple sum 12 weights per voxel and $ 4*(n_x + n_y + n_z) $ for all voxels of the tile. Note that instead of the 12 weights, the 64 combinations of them could be stored instead. That would require 64 * T different weights to be stored.

%	\bibliographystyle{model1-num-names}
	\bibliographystyle{elsarticle-num}
	\bibliography{bsplines_jrnl}

\begin{thebibliography}{10}
\expandafter\ifx\csname url\endcsname\relax
  \def\url#1{\texttt{#1}}\fi
\expandafter\ifx\csname urlprefix\endcsname\relax\def\urlprefix{URL }\fi
\expandafter\ifx\csname href\endcsname\relax
  \def\href#1#2{#2} \def\path#1{#1}\fi

\bibitem{Bartoli2012}
A.~Bartoli, T.~Collins, N.~Bourdel, M.~Canis,
  \href{http://dx.doi.org/10.1016/j.mehy.2012.09.007}{{Computer assisted
  Minimally Invasive Surgery: Is medical Computer Vision the answer to
  improving laparosurgery?}}, Medical Hypotheses 79~(6) (2012) 858--863.
\newblock \href {https://doi.org/10.1016/j.mehy.2012.09.007}
  {\path{doi:10.1016/j.mehy.2012.09.007}}.
\newline\urlprefix\url{http://dx.doi.org/10.1016/j.mehy.2012.09.007}

\bibitem{Bernhardt2017}
S.~Bernhardt, S.~A. Nicolau, L.~Soler, C.~Doignon, {The status of augmented
  reality in laparoscopic surgery as of 2016}, Medical Image Analysis 37 (2017)
  66--90.
\newblock \href {https://doi.org/10.1016/j.media.2017.01.007}
  {\path{doi:10.1016/j.media.2017.01.007}}.

\bibitem{teatini2018assessment}
A.~Teatini, T.~Lang{\o}, B.~Edwin, O.~Elle, et~al., Assessment and comparison
  of target registration accuracy in surgical instrument tracking technologies,
  in: 2018 40th Annual International Conference of the IEEE Engineering in
  Medicine and Biology Society (EMBC), IEEE, 2018, pp. 1845--1848.

\bibitem{sotiras2013deformable}
A.~Sotiras, C.~Davatzikos, N.~Paragios, Deformable medical image registration:
  A survey, IEEE transactions on medical imaging 32~(7) (2013) 1153.

\bibitem{Rueckertetal.1999}
D.~Rueckert, L.~I. Sonoda, C.~Hayes, D.~L. Hill, M.~O. Leach, D.~J. Hawkes,
  {Nonrigid registration using free-form deformations: application to breast MR
  images.}, IEEE Transactions on Medical Imaging 18~(8) (1999) 712--21.
\newblock \href {https://doi.org/10.1109/42.796284}
  {\path{doi:10.1109/42.796284}}.

\bibitem{Ellingwood2016}
N.~D. Ellingwood, Y.~Yin, M.~Smith, C.~L. Lin, {Efficient methods for
  implementation of multi-level nonrigid mass-preserving image registration on
  GPUs and multi-threaded CPUs}, Computer Methods and Programs in Biomedicine
  127 (2016) 290--300.
\newblock \href {https://doi.org/10.1016/J.CMPB.2015.12.018}
  {\path{doi:10.1016/J.CMPB.2015.12.018}}.

\bibitem{Modat2010}
M.~Modat, G.~R. Ridgway, Z.~A. Taylor, M.~Lehmann, J.~Barnes, D.~J. Hawkes,
  N.~C. Fox, S.~Ourselin, {Fast free-form deformation using graphics processing
  units}, Computer Methods and Programs in Biomedicine 98~(3) (2010) 278--284.
\newblock \href {https://doi.org/10.1016/j.cmpb.2009.09.002}
  {\path{doi:10.1016/j.cmpb.2009.09.002}}.

\bibitem{ProgrGuide}
NVIDIA, {CUDA C Programming Guide 9.0} (2017).

\bibitem{smistad2015medical}
E.~Smistad, T.~L. Falch, M.~Bozorgi, A.~C. Elster, F.~Lindseth, Medical image
  segmentation on gpus--a comprehensive review, Medical image analysis 20~(1)
  (2015) 1--18.

\bibitem{gai2013more}
J.~Gai, N.~Obeid, J.~L. Holtrop, X.-L. Wu, F.~Lam, M.~Fu, J.~P. Haldar, W.~H.
  Wen-mei, Z.-P. Liang, B.~P. Sutton, More impatient: A gridding-accelerated
  toeplitz-based strategy for non-cartesian high-resolution 3d mri on gpus,
  Journal of parallel and distributed computing 73~(5) (2013) 686--697.

\bibitem{stone_accelerating_2008}
S.~S. Stone, J.~P. Haldar, S.~C. Tsao, W.-m.~W. Hwu, B.~P. Sutton, Z.-P. Liang,
  \href{http://www.sciencedirect.com/science/article/pii/S0743731508000919}{Accelerating
  advanced {MRI} reconstructions on {GPUs}}, Journal of Parallel and
  Distributed Computing 68~(10) (2008) 1307--1318.
\newblock \href {https://doi.org/10.1016/j.jpdc.2008.05.013}
  {\path{doi:10.1016/j.jpdc.2008.05.013}}.
\newline\urlprefix\url{http://www.sciencedirect.com/science/article/pii/S0743731508000919}

\bibitem{wang2018survey}
H.~Wang, H.~Peng, Y.~Chang, D.~Liang, A survey of gpu-based acceleration
  techniques in mri reconstructions, Quantitative imaging in medicine and
  surgery 8~(2) (2018) 196.

\bibitem{kalaiselvi2017survey}
T.~Kalaiselvi, P.~Sriramakrishnan, K.~Somasundaram, Survey of using gpu cuda
  programming model in medical image analysis, Informatics in Medicine Unlocked
  9 (2017) 133--144.

\bibitem{palomar2018high}
R.~Palomar, J.~G{\'o}mez-Luna, F.~A. Cheikh, J.~Olivares, O.~J. Elle,
  High-performance computation of b{\'e}zier surfaces on parallel and
  heterogeneous platforms, International Journal of Parallel Programming 46~(6)
  (2018) 1035--1062.

\bibitem{nitin2020}
N.~Satpute, R.~Naseem, E.~Pelanis, J.~Gomez-Luna, F.~Alaya~Cheikh, O.~J. Elle,
  J.~Olivares,
  \href{http://www.sciencedirect.com/science/article/pii/S016926071931733X}{Gpu
  acceleration of liver enhancement for tumor segmentation}, Computer Methods
  and Programs in Biomedicine 184 (2020) 105285.
\newblock \href {https://doi.org/https://doi.org/10.1016/j.cmpb.2019.105285}
  {\path{doi:https://doi.org/10.1016/j.cmpb.2019.105285}}.
\newline\urlprefix\url{http://www.sciencedirect.com/science/article/pii/S016926071931733X}

\bibitem{sigg2005fast}
C.~Sigg, M.~Hadwiger, Fast third-order texture filtering, GPU gems 2 (2005)
  313--329.

\bibitem{Ruijters2008}
D.~Ruijters, B.~M. {ter Haar Romeny}, P.~Suetens,
  \href{http://dx.doi.org/10.1080/2151237X.2008.10129269}{{Efficient GPU-Based
  Texture Interpolation using Uniform B-Splines}}, Journal of Graphics, GPU,
  and Game Tools 13~(4) (2008) 61--69.
\newblock \href {https://doi.org/10.1080/2151237X.2008.10129269}
  {\path{doi:10.1080/2151237X.2008.10129269}}.
\newline\urlprefix\url{http://dx.doi.org/10.1080/2151237X.2008.10129269}

\bibitem{Du2016}
X.~Du, J.~Dang, Y.~Wang, S.~Wang, T.~Lei, {A Parallel Nonrigid Registration
  Algorithm Based on B-Spline for Medical Images}, Computational and
  Mathematical Methods in Medicine (2016).
\newblock \href {https://doi.org/10.1155/2016/7419307}
  {\path{doi:10.1155/2016/7419307}}.

\bibitem{Shackleford2010}
J.~A. Shackleford, N.~Kandasamy, G.~C. Sharp, {On developing B-spline
  registration algorithms for multi-core processors}, Physics in Medicine and
  Biology 55~(21) (2010) 6329--6351.
\newblock \href {https://doi.org/10.1088/0031-9155/55/21/001}
  {\path{doi:10.1088/0031-9155/55/21/001}}.

\bibitem{peterlik2018fast}
I.~Peterl{\'\i}k, H.~Courtecuisse, R.~Rohling, P.~Abolmaesumi, C.~Nguan,
  S.~Cotin, S.~Salcudean, Fast elastic registration of soft tissues under large
  deformations, Medical image analysis 45 (2018) 24--40.

\bibitem{lee2015evaluation}
C.~P. Lee, Z.~Xu, R.~P. Burke, R.~Baucom, B.~K. Poulose, R.~G. Abramson, B.~A.
  Landman, Evaluation of five image registration tools for abdominal {CT}:
  Pitfalls and opportunities with soft anatomy, in: Medical Imaging 2015: Image
  Processing, Vol. 9413, International Society for Optics and Photonics, 2015,
  p. 94131N.

\bibitem{Heiselman2018}
J.~S. Heiselman, L.~W. Clements, J.~A. Collins, J.~A. Weis, A.~L. Simpson,
  S.~K. Geevarghese, T.~P. Kingham, W.~R. Jarnagin, M.~I. Miga,
  {Characterization and correction of soft tissue deformation in laparoscopic
  image-guided liver surgery}, Journal of Medical Imaging In Press~(2) (2018).
\newblock \href {https://doi.org/10.1117/1.JMI.5.2.021203}
  {\path{doi:10.1117/1.JMI.5.2.021203}}.

\bibitem{Johnsen2015}
S.~F. Johnsen, S.~Thompson, M.~J. Clarkson, M.~Modat, Y.~Song, J.~Totz,
  K.~Gurusamy, B.~Davidson, Z.~A. Taylor, D.~J. Hawkes, S.~Ourselin,
  \href{https://doi.org/10.1007/978-3-319-24571-3_54}{{Database-Based
  Estimation of Liver Deformation under Pneumoperitoneum for Surgical
  Image-Guidance and Simulation}}, Lecture Notes in Computer Science (including
  subseries Lecture Notes in Artificial Intelligence and Lecture Notes in
  Bioinformatics) 9350 (2015) 450--458.
\newline\urlprefix\url{https://doi.org/10.1007/978-3-319-24571-3_54}

\bibitem{Ruijters2012}
D.~Ruijters, P.~Th{\'{e}}venaz, {GPU prefilter for accurate cubic B-spline
  interpolation}, Computer Journal 55~(1) (2010) 15--20.
\newblock \href {https://doi.org/10.1093/comjnl/bxq086}
  {\path{doi:10.1093/comjnl/bxq086}}.

\bibitem{andersson2016fast}
F.~Andersson, M.~Carlsson, V.~V. Nikitin, Fast algorithms and efficient gpu
  implementations for the radon transform and the back-projection operator
  represented as convolution operators, SIAM Journal on Imaging Sciences 9~(2)
  (2016) 637--664.

\bibitem{carron2017maximum}
J.~Carron, A.~Lewis, Maximum a posteriori cmb lensing reconstruction, Physical
  Review D 96~(6) (2017) 063510.

\bibitem{Volkov2010}
V.~Volkov, {Better performance at lower occupancy}, Proceedings of the GPU
  Technology Conference (2010) 1--75.

\bibitem{Whitehead2011}
N.~Whitehead, A.~Fit-Florea, {Precision {\&} Performance : Floating Point and
  IEEE 754 Compliance for NVIDIA GPUs}, NVIDIA white paper 21~(10) (2011)
  767--75.
\newblock \href {https://doi.org/10.1111/j.1468-2982.2005.00972.x}
  {\path{doi:10.1111/j.1468-2982.2005.00972.x}}.

\bibitem{agnerCPU}
A.~Fog, The microarchitecture of Intel, AMD and VIA CPUs: An optimization guide
  for assembly programmers and compiler makers, Technical University of
  Denmark, 2018th Edition (September 2018).

\bibitem{intelIntrinsics}
Intel, Intel intrinsics guide, software.intel.com, retrieved January 17, 2019
  (2019).

\bibitem{dataset}
O.~Jakob~Elle, A.~Teatini, O.~Zachariadis,
  \href{http://dx.doi.org/10.17632/kj3xcd776k.1}{Data for: {Accelerating}
  {B}-spline {Interpolation} on {GPUs}: {Application} to {Medical} {Image}
  {Registration}}, Mendeley Data (2019).
\newblock \href {https://doi.org/10.17632/kj3xcd776k.1}
  {\path{doi:10.17632/kj3xcd776k.1}}.
\newline\urlprefix\url{http://dx.doi.org/10.17632/kj3xcd776k.1}

\bibitem{Pacioni2015}
A.~Pacioni, M.~Carbone, C.~Freschi, R.~Viglialoro, V.~Ferrari, M.~Ferrari,
  \href{http://dx.doi.org/10.1007/s11548-014-1120-y}{{Patient-specific
  ultrasound liver phantom: materials and fabrication method}}, International
  Journal of Computer Assisted Radiology and Surgery 10~(7) (2015) 1065--1075.
\newblock \href {https://doi.org/10.1007/s11548-014-1120-y}
  {\path{doi:10.1007/s11548-014-1120-y}}.
\newline\urlprefix\url{http://dx.doi.org/10.1007/s11548-014-1120-y}

\bibitem{andrea2018validation}
A.~Teatini, W.~Congcong, P.~Rafael, A.~C. Faouzi, B.~Azeddine, E.~Bj{\o}rn,
  E.~O. Jakob, Validation of stereo vision based liver surface reconstruction
  for image guided surgery, in: Colour and Visual Computing Symposium (CVCS),
  IEEE, 2018, pp. 1--6.

\bibitem{ingenia_MRI}
PHILIPS,
  \href{https://www.theonlinelearningcenter.com/assets/smiles/el_lp_r517/AW9661_UserDoc_HelpTopics_Ingenia_SA/R517_Documentation/en-US/ifu1_p7i_us_pnl.pdf}{{Ingenia:
  Instructions for use}} (2014).
\newline\urlprefix\url{https://www.theonlinelearningcenter.com/assets/smiles/el_lp_r517/AW9661_UserDoc_HelpTopics_Ingenia_SA/R517_Documentation/en-US/ifu1_p7i_us_pnl.pdf}

\bibitem{teatini2019effect}
A.~Teatini, E.~Pelanis, D.~L. Aghayan, R.~P. Kumar, R.~Palomar, {\AA}.~A.
  Fretland, B.~Edwin, O.~J. Elle, The effect of intraoperative imaging on
  surgical navigation for laparoscopic liver resection surgery, Scientific
  Reports 9~(1) (2019) 1--11.

\bibitem{nvidiaTuring2018}
{NVIDIA}, Nvidia {Turing} {Gpu} {Architecture} {Whitepaper} (2018).

\bibitem{ProfGuide}
NVIDIA,
  \href{http://docs.nvidia.com/cuda/profiler-users-guide/index.html}{{Profiler
  User's Guide}}~(September) (2017).
\newline\urlprefix\url{http://docs.nvidia.com/cuda/profiler-users-guide/index.html}

\bibitem{yang2018empirical}
C.~Yang, R.~Gayatri, T.~Kurth, P.~Basu, Z.~Ronaghi, A.~Adetokunbo, B.~Friesen,
  B.~Cook, D.~Doerfler, L.~Oliker, et~al., An empirical roofline methodology
  for quantitatively assessing performance portability, in: 2018 IEEE/ACM
  International Workshop on Performance, Portability and Productivity in HPC
  (P3HPC), IEEE, 2018, pp. 14--23.

\bibitem{amdahl1967validity}
G.~M. Amdahl, Validity of the single processor approach to achieving large
  scale computing capabilities, in: Proceedings of the April 18-20, 1967,
  spring joint computer conference, ACM, 1967, pp. 483--485.

\bibitem{pluim2016truth}
J.~P. Pluim, S.~E. Muenzing, K.~A. Eppenhof, K.~Murphy, The truth is hard to
  make: Validation of medical image registration, in: International Conference
  on Pattern Recognition (ICPR), IEEE, 2016, pp. 2294--2300.

\bibitem{hore2010image}
A.~Hore, D.~Ziou, Image quality metrics: Psnr vs. ssim, in: 2010 20th
  International Conference on Pattern Recognition, IEEE, 2010, pp. 2366--2369.

\bibitem{unser1999splines}
M.~Unser, Splines: A perfect fit for signal and image processing, IEEE Signal
  processing magazine 16~(6) (1999) 22--38.

\bibitem{clements2015validation}
L.~W. Clements, J.~A. Collins, Y.~Wu, A.~L. Simpson, W.~R. Jarnagin, M.~I.
  Miga, Validation of model-based deformation correction in image-guided liver
  surgery via tracked intraoperative ultrasound: preliminary method and
  results, in: Medical Imaging 2015: Image-Guided Procedures, Robotic
  Interventions, and Modeling, Vol. 9415, International Society for Optics and
  Photonics, 2015, p. 94150T.

\bibitem{Kim2012}
H.~Kim, R.~Vuduc, S.~Baghsorkhi, J.~Choi, W.-m. Hwu, {Performance Analysis and
  Tuning for General Purpose Graphics Processing Units (GPGPU)}, Synthesis
  Lectures on Computer Architecture 7 (2012) 1--96.
\newblock \href {https://doi.org/10.2200/S00451ED1V01Y201209CAC020}
  {\path{doi:10.2200/S00451ED1V01Y201209CAC020}}.

\end{thebibliography}
	
\end{document}